# Theoretical examination of QED Hamiltonian and negative-energy orbitals in relativistic molecular orbital theory

Nobuki Inoue[1,a)], Yoshihiro Watanabe[1] and Haruyuki Nakano[1,a)]

The relativistic Hartree–Fock and electron correlation methods without the negative-energy orbital problem are examined on the basis of the quantum electrodynamics (QED) Hamiltonian. First, several QED Hamiltonians previously proposed are sifted by the orbital rotation invariance, the charge conjugation and time reversal invariance, and the nonrelativistic limit. A new total energy expression is then proposed, in which a counter term corresponding to the energy of the polarized vacuum is subtracted from the total energy. This expression prevents the possibility of total energy divergence due to electron correlations, stemming from the fact that the QED Hamiltonian does not conserve the number of particles. Finally, based on the Hamiltonian and energy expression, the Dirac–Hartree–Fock (DHF) and electron correlation methods are reintroduced. The resulting QED-based DHF equation has the same form as the conventional DHF equation, but also formally describes systems specific to QED, such as the virtual positrons in the hydride ion and the positron in positronium. Three electron correlation methods are derived: the QED-based configuration interactions and single- and multireference perturbation methods. Numerical calculations show that the total energy of the QED Hamiltonian indeed diverges and that the counter term is effective in avoiding the divergence. The theoretical examinations in the present article suggest that the molecular orbital (MO) methods based on the QED Hamiltonian not only solve the problem of the negative-energy solutions of the relativistic MO method, but also provide a relativistic formalism to treat systems containing positrons.

## I. INTRODUCTION

The relativistic electronic structure theory, which is based on the Dirac equation, is a powerful tool to understand the properties of systems containing heavy or superheavy elements.[1] Nowadays, the relativistic Hartree–Fock (HF) method[2,3] and electron correlation methods, such as density functional theory,[4] Møller–Plesset (MP) perturbation method,[5] configuration interaction (CI) method,[6] multiconfiguration self-consistent field method,[7] multireference perturbation method,[8–10] multireference coupled-cluster method,[11] two-time Green's function method,[12] and polarization propagator method[13] have been developed and used. The relativistic electronic structure theory is also useful to understand magnetic phenomena because the spin–orbit interaction is naturally included through the Dirac equation. In fact, the nuclear magnetic resonance (NMR) shielding constants can be reproduced by taking into account the spin–orbit interaction.[14] One of the major differences between the Dirac equation and the nonrelativistic Schrödinger equation is that the former has negative kinetic energy solutions as well as positive kinetic energy solutions. This is based on the fact that the relativistic dispersion relation between energy and momentum is a quadratic equation: $E^2 = m^2c^4 + p^2c^2$, i.e., $E = \pm(m^2c^4 + p^2c^2)^{1/2}$. The negative kinetic energy solutions are often problematic in their interpretation and treatment; a negative kinetic energy is unacceptable, and the existence of negative kinetic energy levels can cause collapse of electronic states. This negative kinetic energy is conventionally referred to simply as "negative energy," and we hereafter follow this convention.

[1] *Department of Chemistry, Graduate School of Science, Kyushu University, 744 Motooka, Nishi-ku, Fukuoka 819-0395, Japan*
[a] Authors to whom correspondence should be addressed: ion@ccl.scc.kyushu-u.ac.jp and nakano@chem.kyushu-univ.jp

This problem in the handling of the negative-energy solutions also appears in relativistic molecular orbital (MO) methods when handling excited configurations in the CI method. The problem of the negative-energy solution itself has already been solved by the quantum field theory. However, in the bound state problem of many electrons, it is difficult to select a one-particle state when the field is quantized. Moreover, combining electron correlation and radiation correction with high accuracy in quantum chemical calculations requires high computational cost.

The goal of this study is to construct a framework of MO theory consistent with the quantum field theory for the treatment of negative-energy solutions. Specifically, the conventional CI is reinterpreted from the viewpoint of the quantum field theory.

A simple many-body relativistic Hamiltonian, the Dirac–Coulomb (DC) Hamiltonian, is defined by

$$H^{DC} = \sum_i \left( \begin{bmatrix} \mathbf{0}_2 & c\boldsymbol{\sigma} \cdot \mathbf{p}_i \\ c\boldsymbol{\sigma} \cdot \mathbf{p}_i & -2mc^2\mathbf{1}_2 \end{bmatrix} - \sum_A \frac{Z_A}{|\mathbf{r}_i - \mathbf{R}_A|} \begin{bmatrix} \mathbf{1}_2 & \mathbf{0}_2 \\ \mathbf{0}_2 & \mathbf{1}_2 \end{bmatrix} \right) + \sum_{i>j} \frac{\mathbf{1}_4(i)\mathbf{1}_4(j)}{|\mathbf{r}_i - \mathbf{r}_j|}. \quad (1)$$

The former operator in the first term on the rhs is the relativistic kinetic energy operator. This operator gives all orbitals an energy expectation value larger than 0 or smaller than $-2mc^2$. The orbitals with positive and negative energies are called positive- and negative-energy orbitals, respectively. The positive-energy orbitals can be related with the nonrelativistic electronic orbitals when they are expressed within the nonrelativistic framework, whereas negative-energy orbitals do not correspond to anything without the Dirac sea idea in the nonrelativistic theory. Needless to say, negative energy is unacceptable in reality. It is necessary to consider the form of the second-quantized relativistic Hamiltonian with the existence of the negative-energy orbitals in mind. One approach is to develop a method to treat the negative-energy electrons using negative-energy orbitals. The second-quantized Hamiltonian for this approach is referred to as the virtual pair approximation (VPA) Hamiltonian.[15] Calculations incorporating the VPA Hamiltonian inherently include nonphysical negative-energy solutions. This is a problem because electrons with negative energies have never been observed. Another problem lies in the time evolution of the state that does not include negative-energy electrons, which corresponds to the ground state of the nonrelativistic Hamiltonian. In this case, electrons are allowed to transit to the negative-energy orbital. In other words, the electronic ground state can collapse.

One way to avoid such nonphysical situations is to restrict the system so that it contains only positive-energy orbitals. The second-quantized Hamiltonian so restricted is referred to as the no virtual pair approximation (NVPA) Hamiltonian.[16] This NVPA Hamiltonian does not include the negative-energy orbitals, and the problems observed in VPA do not occur. However, the NVPA has other problems, e.g., the Fock space used for NVPA is not complete. Another problem is that some quantized relativistic Hamiltonians break the causality by removing the negative-energy solutions.[17] Such problems present in the VPA and NVPA Hamiltonians do not appear at the HF level because in the system composed of only electrons, negative-energy orbitals are not occupied and are therefore not included in the ground configuration of the HF method. In other words, the problems occur when electron correlation is considered. In brief, the "correct" relativistic electron correlation theory must include negative-energy orbitals in the Hamiltonian but must not include negative energies in the spectrum. To satisfy these requirements, it is necessary to go beyond the VPA and NVPA Hamiltonians.

An effective way to overcome the problems associated with negative-energy solutions is to employ the quantum electrodynamics (QED) Hamiltonian.[18–22] In the QED Hamiltonian, the relationship between particles and their holes is inverted in negative-energy orbitals. The hole that is treated as a particle is the positron. Therefore, the QED Hamiltonian does not involve a negative kinetic energy problem as the kinetic energy of the positron is inverted to be positive. Several different QED Hamiltonians have been proposed to date,[18–24] depending on the reference vacuum and the type of contraction of the creation/annihilation operators. Because there should be only one "legitimate" Hamiltonian describing nature, we need to examine which one should be adopted through theoretical requirements.

In this article, we discuss the Hamiltonians in terms of orbital rotation invariance, charge conjugation and time reversal (CT) invariance, and the nonrelativistic limit, and propose the relativistic MO theory based on the proposed Hamiltonian.

The arrangement of this article is as follows. In Sec. II, after a review of conventional second-quantized relativistic Hamiltonians, we propose a proper Hamiltonian that satisfies physical requirements, and rederive the relativistic MO theory based on it. In Sec. III, some of the theoretical considerations in Sec. II are confirmed numerically. The conclusions are presented in Sec. IV.

## II. THEORY

In this section, we discuss a relativistic MO theory that does not involve problems of negative-energy solutions. In our discussion, the quantum electromagnetic field is ignored for simplicity—we focus on the problem of negative-energy solutions. This means that we use the DC Hamiltonian instead of the gauge-invariant full QED Hamiltonian or the Dirac–Coulomb–Breit Hamiltonian. Therefore, the QED we are treating in this paper is the so-called "no-photon QED". The other important aspect of the QED, namely, quantum effects of an electromagnetic field,[25–31] is not addressed in this paper. First, we briefly review second-quantized relativistic Hamiltonians based on the DC Hamiltonian—this is necessary for the later discussions. Second, we discuss physical conditions required for the second-quantized relativistic Hamiltonian and propose expressions of the second-quantized relativistic Hamiltonian and total energy satisfying the physical conditions. We then rederive a MO theory based on these expressions.

### A. Brief review of second-quantized relativistic Hamiltonians

For nonrelativistic Hamiltonians, the second-quantized form, written in terms of creation and annihilation operators, is uniquely specified. In contrast, for relativistic Hamiltonians, the second-quantized form is not necessarily uniquely specified because of several possibilities in the treatment of negative-energy orbitals. For readers' convenience, we review some second-quantized forms of the DC Hamiltonians before commencing with the main discussion.

The first is the VPA Hamiltonian,[15]

$$H^{\text{VPA}} = \sum_{pq}^{\text{all}} h_{pq} a_p^\dagger a_q + \frac{1}{2} \sum_{pqrs}^{\text{all}} (pq|rs) a_p^\dagger a_r^\dagger a_s a_q , \qquad (2)$$

where $a_p^\dagger$ and $a_r^\dagger$ are electron creation operators and $a_q$ and $a_s$ are electron annihilation operators. Note that the orbital sums are taken for all the MOs. The integrals $h_{pq}$ and $(pq|rs)$ are defined as

$$h_{pq} = \int d^3\mathbf{r}\, \psi_p^\dagger(\mathbf{r}) \left( \begin{bmatrix} \mathbf{0}_2 & c\boldsymbol{\sigma}\cdot\mathbf{p} \\ c\boldsymbol{\sigma}\cdot\mathbf{p} & -2mc^2 \mathbf{1}_2 \end{bmatrix} - \sum_A \frac{Z_A}{|\mathbf{r}-\mathbf{R}_A|} \begin{bmatrix} \mathbf{1}_2 & \mathbf{0}_2 \\ \mathbf{0}_2 & \mathbf{1}_2 \end{bmatrix} \right) \psi_q(\mathbf{r}) , \qquad (3)$$

$$(pq|rs) = \iint d^3\mathbf{r}_1 d^3\mathbf{r}_2\, \psi_p^\dagger(\mathbf{r}_1) \psi_q(\mathbf{r}_1) \frac{1}{|\mathbf{r}_1-\mathbf{r}_2|} \psi_r^\dagger(\mathbf{r}_2) \psi_s(\mathbf{r}_2) , \qquad (4)$$

and the vacuum state for the VPA Hamiltonian is $|\text{empty}\rangle$. Here, $|\text{empty}\rangle$ is the *completely empty state*; in other words, the state that vanishes by an operation of any (i.e., regardless of positive- or negative-electron energy) electron annihilation operators.

The second is the NVPA Hamiltonian,[16]

$$H^{\text{NVPA}} = \sum_{pq}^{(+)} h_{pq} a_p^\dagger a_q + \frac{1}{2} \sum_{pqrs}^{(+)} (pq|rs) a_p^\dagger a_r^\dagger a_s a_q . \qquad (5)$$

Here, the orbital sums (+) are taken only for the positive-energy MOs. The integrals $h_{pq}$ and $(pq|rs)$ are the same as those in VPA, and the vacuum state of the NVPA Hamiltonian is also $|\text{empty}\rangle$.

The third is the QED Hamiltonian. Several different formulations of the QED Hamiltonian have been proposed, depending on the reference vacuums and other factors. The following Hamiltonian, which we refer to as "QED(MO-CNC)" in the present paper, is a widely known QED Hamiltonian[18–22]

$$\begin{aligned}
H^{\text{QED(MO-CNC)}} &= \sum_{pq}^{(+)} h_{pq} a_p^\dagger a_q - \sum_{pq}^{(-)} h_{pq} b_q^\dagger b_p + \sum_p^{(+)} \sum_q^{(-)} h_{pq} a_p^\dagger b_q^\dagger + \sum_q^{(+)} \sum_p^{(-)} h_{pq} b_p a_q \\
&+ \frac{1}{2} \Bigg( \sum_{pqrs}^{(+)} (pq|rs) a_p^\dagger a_r^\dagger a_s a_q - 2 \sum_{pq}^{(+)} \sum_{rs}^{(-)} (pq|rs) a_p^\dagger b_s^\dagger b_r a_q \\
&+ 2 \sum_{ps}^{(+)} \sum_{qr}^{(-)} (pq|rs) a_p^\dagger b_q^\dagger b_r a_s + \sum_{pqrs}^{(-)} (pq|rs) b_s^\dagger b_q^\dagger b_p b_r \\
&+ 2 \sum_{pqr}^{(+)} \sum_s^{(-)} (pq|rs) a_p^\dagger a_r^\dagger b_s^\dagger a_q - 2 \sum_r^{(+)} \sum_{pqs}^{(-)} (pq|rs) a_r^\dagger b_s^\dagger b_q^\dagger b_p \\
&+ 2 \sum_{pqs}^{(+)} \sum_r^{(-)} (pq|rs) a_p^\dagger b_r a_s a_q - 2 \sum_s^{(+)} \sum_{pqr}^{(-)} (pq|rs) b_q^\dagger b_p b_r a_s \\
&+ \sum_{pr}^{(+)} \sum_{qs}^{(-)} (pq|rs) a_p^\dagger a_r^\dagger b_s^\dagger b_q^\dagger + \sum_{pr}^{(+)} \sum_{qs}^{(-)} (pq|rs) b_p b_r a_s a_q \Bigg),
\end{aligned} \qquad (6)$$

where $b_p^\dagger$ and $b_p$ are creation and annihilation operators of a positron, respectively. Saue and Visscher proposed another QED Hamiltonian[23] that takes the free electron vacuum as a reference,

$$H^{\text{QED(free-CNC)}} = \sum_{pq}^{\text{all}} h_{pq} a_p^\dagger a_q + \frac{1}{2} \sum_{pqrs}^{\text{all}} (pq|rs) a_p^\dagger a_r^\dagger a_s a_q$$
$$- \sum_{p}^{(-)} h_{[pp]} + \frac{1}{2} \sum_{pq}^{(-)} \{([pp]|[qq]) - ([pq]|[qp])\} \quad (7)$$
$$- \sum_{pq}^{\text{all}} \sum_{r}^{(-)} \{(pq|[rr]) - (p[r]|[r]q)\} a_p^\dagger a_q,$$

where the notation using square brackets "[ ]" indicates that the orbitals in them are free-particle orbitals. Hereinafter, we refer to this Hamiltonian Eq. (7) as "QED(free-CNC)". Liu *et al.* also proposed four different forms of QED Hamiltonians[24].

The first form is Eq. (27) in Ref. 24,

$$H^{\text{QED(MO-CCC)}} = \sum_{pq}^{\text{all}} h_{pq} \{a_p^\dagger a_q\}_n + \frac{1}{2} \sum_{pqrs}^{\text{all}} (pq|rs) \{a_p^\dagger a_r^\dagger a_s a_q\}_n$$
$$- \frac{1}{2} \sum_{pq}^{\text{all}} \sum_{r}^{\text{all}} [(pq|rr) - (pr|rq)] \text{sgn}(\varepsilon_r) \{a_p^\dagger a_q\}_n, \quad (8)$$

where the function "sgn" is defined as

$$\text{sgn}(\varepsilon_p) = \begin{cases} 1 & (\psi_p \in \{(+)\}) \\ -1 & (\psi_p \in \{(-)\}) \end{cases}, \quad (9)$$

and $\{\ \}_n$ is the normal-ordered products with respect to the "vacuum configuration" where all negative-energy orbitals are filled with electrons.

The second form is Eq. (98) in Ref. 24,

$$H^{\text{QED(free-CCC)}} = \sum_{pq}^{\text{all}} h_{pq} \{a_p^\dagger a_q\}_{[n]} + \frac{1}{2} \sum_{pqrs}^{\text{all}} (pq|rs) \{a_p^\dagger a_r^\dagger a_s a_q\}_{[n]}$$
$$- \frac{1}{2} \sum_{pq}^{\text{all}} \sum_{r}^{\text{all}} \{(pq|[rr]) - (p[r]|[r]q)\} \text{sgn}(\varepsilon_{[r]}) \{a_p^\dagger a_q\}_{[n]}. \quad (10)$$

The third and fourth forms are

$$H^{\text{QED(free-CCC Mod.1)}} = \sum_{pq}^{\text{all}} h_{pq} \{a_p^\dagger a_q\}_{[n]} + \frac{1}{2} \sum_{pqrs}^{\text{all}} (pq|rs) \{a_p^\dagger a_r^\dagger a_s a_q\}_{[n]}$$
$$- \frac{1}{2} \sum_{pq}^{\text{all}} \sum_{r}^{\text{all}} \{(pq|[rr]) - (p[r]|[r]q)\} \text{sgn}(\varepsilon_{[r]}) \{a_p^\dagger a_q\}_{[n]}$$
$$- \frac{1}{2} \sum_{p}^{\text{all}} \left(h_{[pp]} - h_{[pp]}^{[0]}\right) \text{sgn}(\varepsilon_{[p]}) \quad (11)$$
$$+ \frac{1}{8} \sum_{pq}^{\text{all}} [([pp]|[qq]) - ([pq]|[qp])] \text{sgn}(\varepsilon_{[p]}) \text{sgn}(\varepsilon_{[q]})$$

and

$$H^{\text{QED(free-CCC Mod.2)}} = \sum_{pq}^{\text{all}} h_{pq} \{a_p^\dagger a_q\}_{[n]} + \frac{1}{2} \sum_{pqrs}^{\text{all}} (pq|rs)\{a_p^\dagger a_r^\dagger a_s a_q\}_{[n]}$$
$$- \frac{1}{2} \sum_{pq}^{\text{all}} \sum_{r}^{\text{all}} \{(pq|[rr]) - (p[r]|[r]q)\} \operatorname{sgn}(\varepsilon_{[r]}) \{a_p^\dagger a_q\}_{[n]} \quad (12)$$
$$- \frac{1}{2} \sum_{p}^{\text{all}} (h_{[pp]} - h_{[pp]}^{[0]}) \operatorname{sgn}(\varepsilon_{[p]}),$$

respectively. The previous two Hamiltonians, Eqs. (11) and (12), differ only in constant terms from QED(free-CCC) Eq. (10), hence they are essentially identical. Even considering that these three are essentially identical, at least four QED Hamiltonians are known so far: QED(MO-CNC) Eq. (6), QED(free-CNC) Eq. (7), QED(MO-CCC) Eq. (8), and QED(free-CCC) Eq. (10). Here, the vacuum corresponding to Eq. (6) is $|\text{empty}\rangle$, but the vacuum for Eqs. (7)–(12) is the configuration in which all negative-energy orbitals are filled. This difference of vacuums is not essential, but merely a difference in representation, which will be mentioned in the next subsection, along with the discussion on the QED Hamiltonians.

## B. Systematic derivation of the QED Hamiltonians

In the previous subsection, we noted that there are four candidates for the already known QED Hamiltonians. In this subsection, considering the possibility that there may be other candidates for the QED Hamiltonians, we will reconsider systematic derivation and classification of the QED Hamiltonians. The change from the VPA Hamiltonian to the QED Hamiltonian has been treated in a number of references (Refs. 32–34), but we employ the method described in Refs. 24 and 36. This change is also regarded as the change from the *configuration space* to the *Fock space*.[35] The VPA Hamiltonian can be rewritten as follows:

$$H^{\text{VPA}} = \sum_{pq}^{\text{all}} h_{pq} \{a_p^\dagger a_q\}_n + \frac{1}{2} \sum_{pqrs}^{\text{all}} (pq|rs)\{a_p^\dagger a_r^\dagger a_s a_q\}_n$$
$$+ \sum_{pqrs}^{\text{all}} [(pq|rs) - (ps|rq)] \{a_p^\dagger a_q\}_n \langle 0^{(\text{ref.})}|a_r^\dagger a_s|0^{(\text{ref.})}\rangle$$
$$+ \sum_{pq}^{\text{all}} h_{pq} \langle 0^{(\text{ref.})}|a_p^\dagger a_q|0^{(\text{ref.})}\rangle \quad (13)$$
$$+ \frac{1}{2} \sum_{pqrs}^{\text{all}} [(pq|rs) - (ps|rq)] \langle 0^{(\text{ref.})}|a_p^\dagger a_q|0^{(\text{ref.})}\rangle \langle 0^{(\text{ref.})}|a_r^\dagger a_s|0^{(\text{ref.})}\rangle.$$

Eliminating the constant terms (the fourth and fifth terms in Eq. (13)), we obtain a QED Hamiltonian. To obtain specific forms, it is necessary to specify the reference vacuum $|0^{(\text{ref.})}\rangle$ and furthermore, the contraction $\langle 0^{(\text{ref.})}|a_p^\dagger a_q|0^{(\text{ref.})}\rangle$. Note that the orbitals defining the creation/annihilation operator used in normal ordering $\{\ \}_n$ are automatically determined from the definition of the vacuum.

One choice of reference vacuum is the vacuum where all the negative-energy MOs are filled with electrons:

$$|0^{(\text{ref.})}\rangle \equiv |0_0^{(\text{MO})}\rangle = a_{-1}^\dagger a_{-2}^\dagger \cdots a_{-\infty}^\dagger |\text{empty}\rangle. \quad (14)$$

This vacuum is defined by the MOs[16]; it is referred to as a "floating vacuum" in Ref. 24.

Another choice of the reference vacuum is the vacuum where all the negative-energy free-particle orbitals are filled with electrons

$$\left|0^{(\text{ref.})}\right\rangle \equiv \left|0_0^{(\text{free})}\right\rangle = a_{[-1]}^\dagger a_{[-2]}^\dagger \cdots a_{[-\infty]}^\dagger \left|\text{empty}\right\rangle, \tag{15}$$

or the vacuum where all the negative-energy Furry orbitals are filled with electrons

$$\left|0^{(\text{ref.})}\right\rangle \equiv \left|0_0^{(\text{Furry})}\right\rangle = a_{[\![-1]\!]}^\dagger a_{[\![-2]\!]}^\dagger \cdots a_{[\![-\infty]\!]}^\dagger \left|\text{empty}\right\rangle, \tag{16}$$

where notation with white square brackets "$[\![\ ]\!]$" indicates that the orbitals in them are the Furry orbitals. These vacuums, as in (15) and (16), are defined in terms of fixed orbitals and are referred to as "frozen vacuums" in Ref. 24. Note that if either of these vacuums is adopted, the normal ordering and the contraction must be redefined for the free-particle or Furry orbitals rather than the original MOs (e.g., $p \to [p]$ or $p \to [\![p]\!]$).

There are also several choices for the contraction. One choice is the following:

$$\left\langle 0^{(\text{ref.})} \left| a_p^\dagger a_q \right| 0^{(\text{ref.})} \right\rangle \equiv 0, \tag{17}$$

which we hereafter refer to as the "constantly null contraction (CNC)". Another choice is the charge-conjugate contraction (CCC) as proposed by Liu et al.[22,24,36]

$$\left\langle 0^{(\text{ref.})} \left| a_p^\dagger a_q \right| 0^{(\text{ref.})} \right\rangle \equiv \begin{cases} (-1/2)\delta_{pq} & (p,q \in \{(+)\}) \\ (1/2)\delta_{pq} & (p,q \in \{(-)\}) \\ 0 & (\text{other cases}) \end{cases} \tag{18}$$

and the last one is the conventional contraction (cC)

$$\left\langle 0^{(\text{ref.})} \left| a_p^\dagger a_q \right| 0^{(\text{ref.})} \right\rangle \equiv \begin{cases} \delta_{pq} & (p,q \in \{(-)\}) \\ 0 & (\text{other cases}) \end{cases} \tag{19}$$

Nine different QED Hamiltonians are possible from three different vacuums and three different contractions. If CNC is employed and, in addition, the reference vacuum of $\left|0_0^{(\text{MO})}\right\rangle$, $\left|0_0^{(\text{free})}\right\rangle$, and $\left|0_0^{(\text{Furry})}\right\rangle$ is employed, then the QED(MO-CNC) Hamiltonian

$$\begin{aligned} H^{\text{QED(MO-CNC)}} = &\sum_{pq}^{\text{all}} h_{pq} a_p^\dagger a_q + \frac{1}{2} \sum_{pqrs}^{\text{all}} (pq|rs) a_p^\dagger a_r^\dagger a_s a_q \\ &- \sum_p^{(-)} h_{pp} + \frac{1}{2} \sum_{pq}^{(-)} \left[(pp|qq) - (pq|qp)\right] \\ &- \sum_{pq}^{\text{all}} \sum_r^{(-)} \left[(pq|rr) - (pr|rq)\right] a_p^\dagger a_q \end{aligned} \tag{20}$$

and QED(free-CNC) Hamiltonian (Eq. (7)), and QED(Furry-CNC) Hamiltonian

$$H^{\text{QED(Furry-CNC)}}$$
$$= \sum_{pq}^{\text{all}} h_{pq} a_p^\dagger a_q + \frac{1}{2} \sum_{pqrs}^{\text{all}} (pq|rs) a_p^\dagger a_r^\dagger a_s a_q$$
$$- \sum_p^{(-)} h_{[\![pp]\!]} + \frac{1}{2} \sum_{pq}^{(-)} \{([\![pp]\!]|[\![qq]\!]) - ([\![pq]\!]|[\![qp]\!])\} \qquad (21)$$
$$- \sum_{pq}^{\text{all}} \sum_r^{(-)} \{(pq|[\![rr]\!]) - (p[\![r]\!]|[\![r]\!]q)\} a_p^\dagger a_q$$

are obtained, respectively. Note that Eq. (20) is equivalent to Eq. (6). In fact, Eq. (20) can also be obtained by replacing the operators in Eq. (6) as $b_p^\dagger \to a_p$ and $b_p \to a_p^\dagger$, and ordering them with the anticommutation relations. The vacuum corresponding to Eq. (6) is $|\text{empty}\rangle$ and that corresponding to Eq. (20) is $|0_0^{(\text{MO})}\rangle$, which is the same vacuum, but apparently different.

Then, if CCC is employed and, in addition, the reference vacuum of $|0_0^{(\text{MO})}\rangle$, $|0_0^{(\text{free})}\rangle$, and $|0_0^{(\text{Furry})}\rangle$ is employed, then the QED(MO-CCC) Hamiltonian (Eq. (8))

$$H^{\text{QED(MO-CCC)}} = \sum_{pq}^{\text{all}} h_{pq} a_p^\dagger a_q + \frac{1}{2} \sum_{pqrs}^{\text{all}} (pq|rs) a_p^\dagger a_r^\dagger a_s a_q$$
$$- \sum_p^{(-)} h_{pp} + \frac{1}{2} \sum_p^{(-)} \sum_q^{(+)} [(pp|qq) - (pq|qp)] \qquad (22)$$
$$- \frac{1}{2} \sum_{pq}^{\text{all}} \sum_r^{\text{all}} [(pq|rr) - (pr|rq)] a_p^\dagger a_q,$$

QED(free-CCC) Hamiltonian (Eq. (10))

$$H^{\text{QED(free-CCC)}} = \sum_{pq}^{\text{all}} h_{pq} a_p^\dagger a_q + \frac{1}{2} \sum_{pqrs}^{\text{all}} (pq|rs) a_p^\dagger a_r^\dagger a_s a_q$$
$$- \sum_p^{(-)} h_{[pp]} + \frac{1}{2} \sum_p^{(-)} \sum_q^{(+)} \{([pp]|[qq]) - ([pq]|[qp])\} \qquad (23)$$
$$- \frac{1}{2} \sum_{pq}^{\text{all}} \sum_r^{\text{all}} \{(pq|[rr]) - (p[r]|[r]q)\} a_p^\dagger a_q,$$

and QED(Furry-CCC) Hamiltonian

$$H^{\text{QED(Furry-CCC)}} = \sum_{pq}^{\text{all}} h_{pq} a_p^\dagger a_q + \frac{1}{2} \sum_{pqrs}^{\text{all}} (pq|rs) a_p^\dagger a_r^\dagger a_s a_q$$
$$- \sum_p^{(-)} h_{[\![pp]\!]} + \frac{1}{2} \sum_p^{(-)} \sum_q^{(+)} \{([\![pp]\!]|[\![qq]\!]) - ([\![pq]\!]|[\![qp]\!])\} \qquad (24)$$
$$- \frac{1}{2} \sum_{pq}^{\text{all}} \sum_r^{\text{all}} \{(pq|[\![rr]\!]) - (p[\![r]\!]|[\![r]\!]q)\} a_p^\dagger a_q$$

are obtained, respectively. Here, the equivalence of Eq. (8) and Eq. (22), and Eq. (10) and Eq. (23), can be derived in the same way as before using the anticommutation relations.

Finally, if cC is employed and, in addition the reference vacuum of $|0_0^{(\text{MO})}\rangle$, $|0_0^{(\text{free})}\rangle$,

and $\left|0_0^{(\text{Furry})}\right\rangle$ is employed, then the QED(MO-cC) Hamiltonian

$$H^{\text{QED(MO-cC)}} = \sum_{pq}^{\text{all}} h_{pq} a_p^\dagger a_q + \frac{1}{2} \sum_{pqrs}^{\text{all}} (pq|rs) a_p^\dagger a_r^\dagger a_s a_q \\ - \sum_p^{(-)} h_{pp} - \frac{1}{2} \sum_{pq}^{(-)} \left[ (pp|qq) - (pq|qp) \right],$$  (25)

QED(free-cC) Hamiltonian

$$H^{\text{QED(free-cC)}} = \sum_{pq}^{\text{all}} h_{pq} a_p^\dagger a_q + \frac{1}{2} \sum_{pqrs}^{\text{all}} (pq|rs) a_p^\dagger a_r^\dagger a_s a_q \\ - \sum_p^{(-)} h_{[pp]} - \frac{1}{2} \sum_{pq}^{(-)} \left\{ ([pp]|[qq]) - ([pq]|[qp]) \right\},$$  (26)

and QED(Furry-cC) Hamiltonian

$$H^{\text{QED(Furry-cC)}} = \sum_{pq}^{\text{all}} h_{pq} a_p^\dagger a_q + \frac{1}{2} \sum_{pqrs}^{\text{all}} (pq|rs) a_p^\dagger a_r^\dagger a_s a_q \\ - \sum_p^{(-)} h_{[\![pp]\!]} - \frac{1}{2} \sum_{pq}^{(-)} \left\{ ([\![pp]\!]|[\![qq]\!]) - ([\![pq]\!]|[\![qp]\!]) \right\}$$  (27)

are obtained, respectively.

We need to choose the most proper second-quantized form of the DC Hamiltonian from these Hamiltonians. The discussion that follows requires some conditions that seem physically appropriate for many-body systems of electrons and positrons.

## C. Fundamental requirements for the QED Hamiltonian: orbital rotation invariance, CT invariance, and nonrelativistic limit

In this section, we discuss the orbital rotation invariance, CT invariance, and nonrelativistic limit of the Hamiltonian as requirements for the second-quantized relativistic Hamiltonian. We focus on these three requirements because the orbital rotation invariance is closely related to the partitioning of the orbital space describing electrons and positrons, the CT invariance is related to the symmetry between electrons and positrons and, in the nonrelativistic limit, the symmetry between electrons and positrons must also hold. After sifting the candidates for the Hamiltonian using these three requirements, we discuss the expression of total energy for the Hamiltonian.

### 1. Orbital rotation invariance

First, we discuss the orbital rotation invariance of the Hamiltonian. Orbital rotation including negative-energy orbitals was first considered in Ref. 37. Consider the orbital rotation, namely, the unitary transformation for MOs:

$$\varphi_i \to \varphi_i' = \sum_n U_{ni} \varphi_n .$$  (28)

The corresponding unitary transformation for the annihilation operators is given by

$$a_i \to a_i' = \sum_n \left(U^\dagger\right)_{in} a_n .$$  (29)

Using these expressions, the one- and two-electron terms of the VPA Hamiltonian are transformed as follows:

$$\sum_{ij}^{\text{all}} h_{ij} a_i^\dagger a_j \to \sum_{ij}^{\text{all}} h'_{ij} a_i'^\dagger a_j' = \sum_{ijmn}^{\text{all}} (U^\dagger)_{im} h_{mn} U_{nj} U_{mi} a_m^\dagger (U^\dagger)_{jn} a_n \quad (30)$$

$$= \sum_{mn}^{\text{all}} h_{mn} a_m^\dagger a_n,$$

$$\sum_{ijkl}^{\text{all}} (ij|kl) a_i^\dagger a_k^\dagger a_l a_j \to \sum_{ijkl}^{\text{all}} (ij|kl)' a_i'^\dagger a_k'^\dagger a_l' a_j'$$

$$= \sum_{ijklpqrs}^{\text{all}} (U^\dagger)_{ip} U_{qj} (pq|rs)(U^\dagger)_{kr} U_{sl} U_{pi} a_p^\dagger U_{rk} a_r^\dagger (U^\dagger)_{ls} a_s (U^\dagger)_{jq} a_q \quad (31)$$

$$= \sum_{pqrs}^{\text{all}} (pq|rs) a_p^\dagger a_r^\dagger a_s a_q.$$

These results show that the VPA Hamiltonian is invariant for the orbital rotation. In the same manner, we can easily show that the QED(free-CNC), QED(Furry-CNC), QED(free-CCC), QED(Furry-CCC), QED(free-cC), and QED(Furry-cC) Hamiltonians are also invariant for orbital rotation.

In contrast, the QED(MO-CNC) Hamiltonian is not invariant for the transformation. To see how the QED Hamiltonian is transformed, we take a term $\sum_{ij}^{\text{all}} \sum_{k}^{(-)} [(ij|kk) - (ik|kj)] a_i^\dagger a_j$ in $H^{\text{QED(MO-CNC)}}$ as an example. This term is transformed by the orbital rotation as follows:

$$\sum_{ij}^{\text{all}} \sum_{k}^{(-)} [(ij|kk) - (ik|kj)] a_i^\dagger a_j \to \sum_{ijrs}^{\text{all}} \sum_{k}^{(-)} [(ij|kk)' - (ik|kj)'] a_i'^\dagger a_j'$$

$$= \sum_{ijpqrs}^{\text{all}} \sum_{k}^{(-)} (U^\dagger)_{ip} U_{qj} [(pq|rs) - (ps|rq)] (U^\dagger)_{kr} U_{sk} U_{pi} a_p^\dagger (U^\dagger)_{jq} a_q \quad (32)$$

$$= \sum_{pqrs}^{\text{all}} \sum_{k}^{(-)} (U^\dagger)_{kr} U_{sk} [(pq|rs) - (ps|rq)] a_p^\dagger a_q$$

$$\neq \sum_{pq}^{\text{all}} \sum_{r}^{(-)} [(pq|rr) - (pr|rq)] a_p^\dagger a_q.$$

This is clearly not invariant because the summation for $k$ is taken not for all the orbitals, but only for the negative-energy orbitals. The same is true for the other terms, showing that the total QED Hamiltonian is not orbital rotation invariant. In other words, QED(MO-CNC), QED(MO-CCC), and QED(MO-cC) Hamiltonians are orbital rotation invariant only for special orbital rotations that do not mix positive- and negative-energy orbitals, i.e., orbital rotations represented by block diagonal matrices for the positive- and negative-energy orbital parts. The general orbital rotations obviously have no such special properties and consequently change the QED(MO-CNC), QED(MO-CCC), and QED(MO-cC) Hamiltonians. This means that even if we constitute the wave function using all possible configurations, namely, excitations, pair creations, and pair annihilations, we cannot obtain unique energy spectra from the Hamiltonian. This is in sharp contrast to the nonrelativistic Hamiltonian, for which we can obtain the unique spectra using the full CI for any orbital sets made from a common basis set. These situations are also the case for the NVPA Hamiltonian.

The discussion thus far has shown that some QED Hamiltonians are not orbital rotation invariant. This fact seems to indicate that some QED Hamiltonians have a problem of giving different spectra depending on the orbital sets. However, this problem is resolved by considering the time dependency of the states. As the spectra are the expectation values of the total energy in the stationary state, the orbitals used to calculate the spectra must be ones capable of constituting the stationary state, and such orbitals must themselves be stationary: i.e., they must not evolve in time. In other words, by using the stationary orbitals, the spectra are uniquely determined even without orbital rotation invariance in stationary states, which is in contrast to general (nonstationary) states, where any orbitals are allowed. Therefore, the orbital rotation invariance for the QED Hamiltonians itself is not necessarily essential. Note here that in this paper the term "stationary" is used in two senses: one is used to mean that a variable does not evolve in time, the other is used to mean that the derivative of a function or functional derivative is zero (i.e., at the maximum, minimum, or saddle point). We write simply "stationary" when we use it in the former sense and "stationary point" in the latter sense.

Then, how can the stationary orbitals that are suitable for the QED-level MO calculations be obtained? The answer to this question lies in finding the MOs giving a stationary point of total energy, as we usually do in the nonrelativistic electronic structure calculations. In fact, from the time-dependent variational principle, the following shows that MOs giving a stationary point of total energy do not evolve in time.

Let $|\Psi(t)\rangle$ be the wave function of a state, and express its time dependence as a linear combination of determinants $|\Phi_I(t)\rangle$

$$|\Psi(t)\rangle = \sum_I C_I(t)|\Phi_I(t)\rangle. \qquad (33)$$

The time dependence of the determinants $|\Phi_I(t)\rangle$ is further expressed using time-dependent MO coefficients $q_{mi}(t)$ for the MOs $\varphi_i(t)$

$$\varphi_i(t) = \sum_m q_{mi}(t)\varphi_m(0) \qquad (34)$$

and creation operators $a_m^\dagger(0)$ at $t=0$ as

$$|\Phi_I(t)\rangle = \prod_i^{\text{occ.}I}\left(\sum_m q_{mi}(t)\, a_m^\dagger(0)\right)|\text{empty}\rangle, \qquad (35)$$

where $i$ in the multiplication runs over the occupied MOs in determinant $I$. To describe the time evolution of the state $|\Psi(t)\rangle$, we define the following Lagrangian:

$$\begin{aligned}L = &\frac{i}{2}\langle\Psi(t)|\dot\Psi(t)\rangle - \frac{i}{2}\langle\dot\Psi(t)|\Psi(t)\rangle - \langle\Psi(t)|H|\Psi(t)\rangle \\ &- \sum_{ij}\lambda_{ij}\left(\frac{\mathrm{d}}{\mathrm{d}t}\sum_k\left(q_{ki}^*(t)q_{kj}(t)\right)\right) - \Lambda\left(\frac{\mathrm{d}}{\mathrm{d}t}\sum_I\left(C_I^*(t)C_I(t)\right)\right),\end{aligned} \qquad (36)$$

where the "over-dot" means the time derivative, as usual. Because the ket $|\Psi(t)\rangle$ depends only on $q_{pq}$ and $C_I$, but not on their complex conjugate $q_{pq}^*$ and $C_I^*$, and vice versa for the bra $\langle\Psi(t)|$, the time derivative of the state $|\dot\Psi(t)\rangle$ is written as

$$\left|\dot{\Psi}(t)\right\rangle = \sum_{pq}\left|\frac{\partial \Psi}{\partial q_{pq}}(t)\right\rangle \dot{q}_{pq} + \sum_{I}\left|\frac{\partial \Psi}{\partial C_{I}}(t)\right\rangle \dot{C}_{I},$$

$$\left\langle\dot{\Psi}(t)\right| = \sum_{pq}\left\langle\frac{\partial \Psi}{\partial q_{pq}}(t)\right| \dot{q}_{pq}^{*} + \sum_{I}\left\langle\frac{\partial \Psi}{\partial C_{I}}(t)\right| \dot{C}_{I}^{*}. \tag{37}$$

Then, applying the Euler–Lagrange equation to the Lagrangian (36) gives the equation of motions for the MOs

$$\frac{\partial L}{\partial q_{pq}^{*}} - \frac{d}{dt}\frac{\partial L}{\partial \dot{q}_{pq}^{*}}$$
$$= i\sum_{rs} A_{pq,rs}\,\dot{q}_{rs}(t) + i\sum_{I} a_{I,pq}^{*} \dot{C}_{I}(t) - \frac{\partial E}{\partial q_{pq}^{*}} \tag{38}$$
$$= 0,$$

as well as that for CI coefficients

$$\frac{\partial L}{\partial C_{I}^{*}} - \frac{d}{dt}\frac{\partial L}{\partial \dot{C}_{I}^{*}}$$
$$= i\sum_{rs} a_{I,rs}\,\dot{q}_{rs}(t) + i\dot{C}_{I}(t) - \frac{\partial E}{\partial C_{I}^{*}} \tag{39}$$
$$= 0,$$

where we have used the following as shorthand notations:

$$E \equiv \left\langle \Psi(t)|H|\Psi(t) \right\rangle,$$

$$A_{pq,rs} \equiv \left\langle \frac{\partial \Psi}{\partial q_{pq}}(t)\bigg|\frac{\partial \Psi}{\partial q_{rs}}(t) \right\rangle, \tag{40}$$

$$a_{I,pq} \equiv \left\langle \frac{\partial \Psi}{\partial C_{I}}(t)\bigg|\frac{\partial \Psi}{\partial q_{pq}}(t) \right\rangle.$$

By combining Eqs. (38) and (39) into a matrix representation, we have

$$\begin{bmatrix} \mathbf{A} & \mathbf{a}^{\dagger} \\ \mathbf{a} & \mathbf{1} \end{bmatrix}\begin{bmatrix} \dot{\mathbf{q}} \\ \dot{\mathbf{C}} \end{bmatrix} = \begin{bmatrix} \partial E/\partial \mathbf{q}^{*} \\ \partial E/\partial \mathbf{C}^{*} \end{bmatrix}, \tag{41}$$

which shows that if $\partial E/\partial q_{pq}^{*} = 0$ and $\partial E/\partial C_{I}^{*} = 0$ then $\dot{q}_{rs}(t) = 0$ because the inverse of the coefficient matrix of Eq. (41) exists under usual conditions. In other words, if $E$ is at a stationary point with respect to changes in MO coefficients $q_{pq}(t)$ and CI coefficients $C_{I}(t)$, the orbital coefficients $q_{pq}(t)$ do not evolve in time, which is exactly what we were trying to show.

As shown above, the MOs giving a stationary point of total energy are suitable for stationary state calculations in that they do not evolve in time, beyond the significance of simply giving an optimal solution in the variational method. Consequently, the problem of the QED Hamiltonian not being orbital rotation invariant is solved by using these optimal MOs.

Thus, we have determined the orbital rotation invariance/noninvariance of the Hamiltonians. By employing the proper MOs, even orbital rotation invariant Hamiltonians are not sifted out, but it should be noted that it is not possible to choose arbitrary MOs when using such Hamiltonians for stationary state calculations.

## 2. CT invariance

We next discuss the CT invariance. The CT transformation is a composite transformation of the charge conjugation[22] and time reversal[38] (C and T) transformations, defined as

$$\hat{C}\hat{K} = \hat{K}\hat{C} = \begin{bmatrix} 0 & \mathbf{1}_2 \\ -\mathbf{1}_2 & 0 \end{bmatrix} \quad (42)$$

in matrix form, where $\hat{C}$ and $\hat{K}$ are the C and T transformation operators, respectively. Applying the CT transformation (Eq. (42)) to the Dirac equation

$$\begin{bmatrix} V\mathbf{1}_2 & c\boldsymbol{\sigma}\cdot\mathbf{p} \\ c\boldsymbol{\sigma}\cdot\mathbf{p} & V\mathbf{1}_2 - 2mc^2\mathbf{1}_2 \end{bmatrix} \begin{bmatrix} \Psi^L \\ \Psi^S \end{bmatrix} = E \begin{bmatrix} \Psi^L \\ \Psi^S \end{bmatrix}, \quad (43)$$

we obtain the CT-transformed Dirac equation

$$\begin{bmatrix} -V\mathbf{1}_2 & c\boldsymbol{\sigma}\cdot\mathbf{p} \\ c\boldsymbol{\sigma}\cdot\mathbf{p} & -V\mathbf{1}_2 - 2mc^2\mathbf{1}_2 \end{bmatrix} \begin{bmatrix} \Psi^S \\ -\Psi^L \end{bmatrix} = \left(-E - 2mc^2\right) \begin{bmatrix} \Psi^S \\ -\Psi^L \end{bmatrix} = \tilde{E} \begin{bmatrix} \Psi^S \\ -\Psi^L \end{bmatrix}, \quad (44)$$

where $\tilde{E} = -E - 2mc^2$ is *reduced energy*: if $E$ is negative, $\tilde{E}$ gives positive (normal) energy. This CT-transformed Dirac equation (Eq. (44)) contains a potential of inverse sign that gives antiparticle solutions,[39,40] and the associated eigen spinor has the form of exchanged large and small components. The *reduced energy* corresponds to the energy of positrons without the rest mass energy. Therefore, negative-energy orbitals with negative and positive *reduced energy* represent the bound and unbound positrons, respectively. Here, the fact that the negative-energy solutions are associated with antiparticles by $\tilde{E}$ rather than $E$ indicates that the particle–hole relationship is inverted in the negative-energy solutions.

Based on the properties of the CT transformation for the one-particle Dirac equation, we next discuss the CT transformation for the multiparticle (DC) Hamiltonian. Here, as preparation before CT transformation, we rewrite Eq. (20). Because the indices in the sums run over all the orbitals, including both Kramers pairs ($p$ and $\bar{p}$), we can rewrite Eq. (20) by changing the order of the orbitals in the sums

$$\begin{aligned}
H^{\text{QED(MO-CNC)}} &= \sum_{pq}^{\text{all}} h_{\overline{pq}} a_{\bar{p}}^\dagger a_{\bar{q}} + \frac{1}{2} \sum_{pqrs}^{\text{all}} \left(\overline{pq}\,|\,\overline{rs}\right) a_{\bar{p}}^\dagger a_{\bar{r}}^\dagger a_{\bar{s}} a_{\bar{q}} \\
&\quad - \sum_{p}^{(-)} h_{\overline{pp}} + \frac{1}{2} \sum_{pq}^{(-)} \left[\left(\overline{pp}\,|\,\overline{qq}\right) - \left(\overline{pq}\,|\,\overline{qp}\right)\right] \\
&\quad - \sum_{pq}^{\text{all}} \sum_{r}^{(-)} \left[\left(\overline{pq}\,|\,\overline{rr}\right) - \left(\overline{pr}\,|\,\overline{rq}\right)\right] a_{\bar{p}}^\dagger a_{\bar{q}}.
\end{aligned} \quad (45)$$

It should be noted that it is simply a change in the order of the sum. This manipulation is done to make it easier to see the final result of the CT transformation. Applying CT transformation to the Hamiltonian, the creation/annihilation operators and molecular integrals change as follows:

$$a_p^\dagger \to a_{\bar{p}}, \quad a_p \to a_{\bar{p}}^\dagger, \quad h_{pq} \to -h_{pq}^{\text{CT}} \text{ and } (pq\,|\,rs) \to (pq\,|\,rs)^{\text{CT}}, \quad (46)$$

and we obtain

$$\left(H^{\text{QED(MO-CNC)}}\right)^{\text{CT}} = \sum_{pq}^{\text{all}} h_{pq}^{\text{CT}} a_p^\dagger a_q + \frac{1}{2} \sum_{pqrs}^{\text{all}} (pq|rs)^{\text{CT}} a_p^\dagger a_r^\dagger a_s a_q$$

$$-\sum_p^{(+)} h_{pp}^{\text{CT}} + \frac{1}{2} \sum_{pq}^{(+)} \left[(pp|qq)^{\text{CT}} - (pq|qp)^{\text{CT}}\right] \quad (47)$$

$$-\sum_{pq}^{\text{all}} \sum_r^{(+)} \left[(pq|rr)^{\text{CT}} - (pr|rq)^{\text{CT}}\right] a_p^\dagger a_q$$

with

$$h_{pq}^{\text{CT}} = \int d^3 \mathbf{r}\, \psi_p^\dagger(\mathbf{r}) \begin{bmatrix} -V\mathbf{1}_2 + 2mc^2 \mathbf{1}_2 & c\boldsymbol{\sigma}\cdot\mathbf{p} \\ c\boldsymbol{\sigma}\cdot\mathbf{p} & -V\mathbf{1}_2 \end{bmatrix} \psi_q(\mathbf{r}),$$

$$(pq|rs)^{\text{CT}} = (pq|rs). \quad (48)$$

This CT-transformed Hamiltonian includes the external field potential of inverse sign, and the rest mass energy is attached to the particles, not the antiparticles. This means that the CT-transformed Hamiltonian is essentially the same as if it were formulated by simply swapping the particles and antiparticles. Hence, it gives the same solutions as those of the original Hamiltonian. In this meaning, the QED(MO-CNC) Hamiltonian is CT invariant. In a similar way, QED(free-CNC), QED(Furry-CNC), QED(MO-CCC), QED(free-CCC), and QED(Furry-CCC) Hamiltonians are shown to be CT invariant, as follows:

$$\left(H^{\text{QED(free-CNC)}}\right)^{\text{CT}} = \sum_{pq}^{\text{all}} h_{qp}^{\text{CT}} a_q^\dagger a_p + \frac{1}{2} \sum_{pqrs}^{\text{all}} (pq|rs) a_p^\dagger a_r^\dagger a_s a_q$$

$$-\sum_p^{(+)} h_{[pp]}^{\text{CT}} + \frac{1}{2} \sum_{pq}^{(+)} \left\{([pp]|[qq]) - ([pq]|[qp])\right\} \quad (49)$$

$$-\sum_{pq}^{\text{all}} \sum_r^{(+)} \left\{(pq|[rr]) - (p[r]|[r]q)\right\} a_p^\dagger a_q,$$

$$\left(H^{\text{QED(Furry-CNC)}}\right)^{\text{CT}} = \sum_{pq}^{\text{all}} h_{qp}^{\text{CT}} a_q^\dagger a_p + \frac{1}{2} \sum_{pqrs}^{\text{all}} (pq|rs) a_p^\dagger a_r^\dagger a_s a_q$$

$$-\sum_p^{(+)} h_{[\![pp]\!]}^{\text{CT}} + \frac{1}{2} \sum_{pq}^{(+)} \left\{([\![pp]\!]|[\![qq]\!]) - ([\![pq]\!]|[\![qp]\!])\right\} \quad (50)$$

$$-\sum_{pq}^{\text{all}} \sum_r^{(+)} \left\{(pq|[\![rr]\!]) - (p[\![r]\!]|[\![r]\!]q)\right\} a_p^\dagger a_q,$$

$$\left(H^{\text{QED(MO-CCC)}}\right)^{\text{CT}} = \sum_{pq}^{\text{all}} h_{pq}^{\text{CT}} a_p^\dagger a_q + \frac{1}{2} \sum_{pqrs}^{\text{all}} (pq|rs) a_p^\dagger a_r^\dagger a_s a_q$$

$$-\sum_p^{(+)} h_{pp}^{\text{CT}} + \frac{1}{2} \sum_p^{(+)} \sum_q^{(-)} \left\{(pp|qq) - (pq|qp)\right\} \quad (51)$$

$$-\frac{1}{2} \sum_{pq}^{\text{all}} \sum_r^{\text{all}} \left\{(pq|rr) - (pr|rq)\right\} a_p^\dagger a_q,$$

$$\left(H^{\text{QED(free-CCC)}}\right)^{\text{CT}} = \sum_{pq}^{\text{all}} h_{pq}^{\text{CT}} a_p^\dagger a_q + \frac{1}{2}\sum_{pqrs}^{\text{all}}(pq|rs)a_p^\dagger a_r^\dagger a_s a_q$$

$$-\sum_p^{(+)} h_{[pp]}^{\text{CT}} + \frac{1}{2}\sum_p^{(+)}\sum_q^{(-)}\left\{\left([pp]|[qq]\right)-\left([pq]|[qp]\right)\right\} \quad (52)$$

$$-\frac{1}{2}\sum_{pq}^{\text{all}}\sum_r^{\text{all}}\left\{\left(pq|[rr]\right)-\left(p[r]|[r]q\right)\right\}a_p^\dagger a_q,$$

$$\left(H^{\text{QED(Furry-CCC)}}\right)^{\text{CT}} = \sum_{pq}^{\text{all}} h_{pq}^{\text{CT}} a_p^\dagger a_q + \frac{1}{2}\sum_{pqrs}^{\text{all}}(pq|rs)a_p^\dagger a_r^\dagger a_s a_q$$

$$-\sum_p^{(+)} h_{\llbracket pp \rrbracket}^{\text{CT}} + \frac{1}{2}\sum_p^{(+)}\sum_q^{(-)}\left\{\left(\llbracket pp \rrbracket|\llbracket qq \rrbracket\right)-\left(\llbracket pq \rrbracket|\llbracket qp \rrbracket\right)\right\} \quad (53)$$

$$-\frac{1}{2}\sum_{pq}^{\text{all}}\sum_r^{\text{all}}\left\{\left(pq|\llbracket rr \rrbracket\right)-\left(p\llbracket r \rrbracket|\llbracket r \rrbracket q\right)\right\}a_p^\dagger a_q.$$

In contrast, the VPA, NVPA, QED(MO-cC), QED(free-cC), and QED(Furry-cC) Hamiltonians are not CT invariant—they change forms by CT transformation, beyond the mere exchange of particles and antiparticles. In fact, the CT-transformed VPA, NVPA, and QED(MO-cC), QED(free-cC), and QED(Furry-cC) Hamiltonians are given as

$$\left(H^{\text{VPA}}\right)^{\text{CT}} = \sum_{pq}^{\text{all}} h_{qp}^{\text{CT}} a_q^\dagger a_p + \frac{1}{2}\sum_{pqrs}^{\text{all}}(pq|rs)a_p^\dagger a_r^\dagger a_s a_q$$

$$-\sum_p^{\text{all}} h_{pp}^{\text{CT}} - \sum_{pqr}^{\text{all}}\left[(pq|rr)-(pr|rq)\right]a_p^\dagger a_q + \frac{1}{2}\sum_{pq}^{\text{all}}\left[(pp|qq)-(pq|qp)\right], \quad (54)$$

$$\left(H^{\text{NVPA}}\right)^{\text{CT}} = \sum_{pq}^{(+)} h_{qp}^{\text{CT}} a_q^\dagger a_p + \frac{1}{2}\sum_{pqrs}^{(+)}(pq|rs)a_p^\dagger a_r^\dagger a_s a_q$$

$$-\sum_p^{(+)} h_{pp}^{\text{CT}} - \sum_{pqr}^{(+)}\left[(pq|rr)-(pr|rq)\right]a_p^\dagger a_q + \frac{1}{2}\sum_{pq}^{(+)}\left[(pp|qq)-(pq|qp)\right], \quad (55)$$

$$\left(H^{\text{QED(MO-cC)}}\right)^{\text{CT}} = \sum_{pq}^{\text{all}} h_{pq}^{\text{CT}} a_p^\dagger a_q + \frac{1}{2}\sum_{pqrs}^{\text{all}}(pq|rs)a_p^\dagger a_r^\dagger a_s a_q$$

$$-\sum_p^{(+)} h_{pp}^{\text{CT}} - \frac{1}{2}\sum_{pq}^{(+)}\left\{(pp|qq)-(pq|qp)\right\}$$

$$-\sum_{pqr}^{\text{all}}\left[(pq|rr)-(pr|rq)\right]a_p^\dagger a_q \quad (56)$$

$$+\sum_{pq}^{(+)}\left\{(pp|qq)-(pq|qp)\right\}$$

$$+\frac{1}{2}\sum_p^{(+)}\sum_q^{(-)}\left\{(pp|qq)-(pq|qp)\right\},$$

$$\left(H^{\text{QED(free-cC)}}\right)^{\text{CT}} = \sum_{pq}^{\text{all}} h_{pq}^{\text{CT}} a_p^\dagger a_q + \frac{1}{2}\sum_{pqrs}^{\text{all}} (pq|rs) a_p^\dagger a_r^\dagger a_s a_q$$
$$-\sum_p^{(+)} h_{[pp]}^{\text{CT}} - \frac{1}{2}\sum_{pq}^{(+)} \{([pp]|[qq]) - ([pq]|[qp])\}$$
$$-\sum_{pqr}^{\text{all}} \left[(pq|rr) - (pr|rq)\right] a_p^\dagger a_q \qquad (57)$$
$$+\sum_{pq}^{(+)} \{([pp]|[qq]) - ([pq]|[qp])\}$$
$$+\frac{1}{2}\sum_p^{(+)}\sum_q^{(-)} \{([pp]|[qq]) - ([pq]|[qp])\},$$

and

$$\left(H^{\text{QED(Furry-cC)}}\right)^{\text{CT}} = \sum_{pq}^{\text{all}} h_{pq}^{\text{CT}} a_p^\dagger a_q + \frac{1}{2}\sum_{pqrs}^{\text{all}} (pq|rs) a_p^\dagger a_r^\dagger a_s a_q$$
$$-\sum_p^{(+)} h_{[\![pp]\!]}^{\text{CT}} - \frac{1}{2}\sum_{pq}^{(+)} \{([\![pp]\!]|[\![qq]\!]) - ([\![pq]\!]|[\![qp]\!])\}$$
$$-\sum_{pqr}^{\text{all}} \left[(pq|rr) - (pr|rq)\right] a_p^\dagger a_q \qquad (58)$$
$$+\sum_{pq}^{(+)} \{([\![pp]\!]|[\![qq]\!]) - ([\![pq]\!]|[\![qp]\!])\}$$
$$+\frac{1}{2}\sum_p^{(+)}\sum_q^{(-)} \{([\![pp]\!]|[\![qq]\!]) - ([\![pq]\!]|[\![qp]\!])\},$$

respectively. In the Hamiltonians (54)–(58), new terms have been generated by the CT transformation. Of these new terms, the constant terms are permitted as they only shift the energy spectra, but the one-electron operator terms clearly violate the CT invariance. Thus, of the eleven second-quantized relativistic Hamiltonians, QED(MO-CNC), QED(free-CNC), QED(Furry-CNC), QED(MO-CCC), QED(free-CCC), and QED(Furry-CCC) are CT invariant.

From the discussion on the CT invariance for the Hamiltonians, we can expect that the negative-energy virtual orbitals describe the distribution of a positron, and that by using an appropriate basis set we can describe the positronic orbitals in systems including positrons using negative-energy orbitals. In order to verify these predictions, the exact solutions for the hydrogen-like ion need to be examined in detail. The exact solution of the Dirac equation for the hydrogen-like ion is written in the following form:[19]

$$\psi = \frac{1}{r}\begin{bmatrix} P_{n\kappa}(r)\xi_{\kappa m}(\theta,\phi) \\ iQ_{n\kappa}(r)\xi_{-\kappa m}(\theta,\phi) \end{bmatrix}. \qquad (59)$$

Applying the CT transformation to Eq. (59) yields a spinor with the large and small components interchanged in Eq. (59)

$$\psi^{\text{CT}} = \frac{1}{r}\begin{bmatrix} iQ_{n\kappa}(r)\xi_{-\kappa m}(\theta,\phi) \\ -P_{n\kappa}(r)\xi_{\kappa m}(\theta,\phi) \end{bmatrix}. \qquad (60)$$

Ignoring the difference in radial functions and focusing only on angular symmetry, this indicates that the CT transformation is equivalent to inverting the sign of $\kappa = (l-j)(2j+1)$. In other words, whether or not a negative solution represents a bound state, spinors with angular symmetry of the positronic $s_{1/2}$ orbital have the same angular symmetry as spinors of the electronic $p_{1/2}$ orbital. Such a correspondence on the angular symmetry holds between the spinors with the same $j$: $s_{1/2} \rightleftarrows p_{1/2}$, $p_{3/2} \rightleftarrows d_{3/2}$, $d_{5/2} \rightleftarrows f_{5/2}$, $f_{7/2} \rightleftarrows g_{7/2}$, … In this way, we can relate the function form of the small component to that of the nonrelativistic orbital and thus determine the angular symmetry of the negative-energy solutions as particles. More proactively, we can explicitly represent the positronic orbitals with the basis functions that have angular symmetry with the sign of $\kappa$ reversed from that of the electrons. In this case, one might think that there is still a concern that the extra $r^2$ term in the positive $\kappa$ case included in the following small component:

$$\boldsymbol{\sigma}\cdot\mathbf{p}\, r^l e^{-\zeta r^2}\xi_{\kappa m} \propto \begin{cases} r^{l+1}e^{-\zeta r^2}\xi_{-\kappa m} & \text{(negative } \kappa\text{)} \\ r^{l-1}\left(1-\frac{\zeta}{l+1/2}r^2\right)e^{-\zeta r^2}\xi_{-\kappa m} & \text{(positive } \kappa\text{)}, \end{cases} \quad (61)$$

which is obtained by imposing strict kinetic balance in the Gauss basis functions, may break the CT symmetry and interfere with the description of the positronic orbitals. This concern is not an essential problem, however, because remedies are possible; e.g., (i) using a sufficiently rich basis set, (ii) using *restricted kinetic balance* (RKB)[41–43] for negative $\kappa$ orbitals and *inverse kinetic balance* (IKB)[44] for positive $\kappa$ orbitals as in Ref. 45, and (iii) using *dual kinetic balance* (DKB).[46] The important item to note here is that if the Hamiltonian is CT invariant and an appropriate basis set is employed, the negative-energy orbitals are interpreted consistently as representing the distribution of a positron.

## 3. Nonrelativistic limit of QED Hamiltonians

We now discuss the nonrelativistic limit of the QED Hamiltonians. The QED Hamiltonian solves the problem of negative-energy solutions by considering the electrons in the negative-energy solution as positron holes. Therefore, the proper QED Hamiltonian must be capable of describing electron–positron many-body systems. In particular, the nonrelativistic limit of the QED Hamiltonian must coincide with an appropriate nonrelativistic Hamiltonian, except for the difference in the constant terms. Based on this requirement, we verify the nonrelativistic limit of the QED Hamiltonians.

For the description of electron–positron systems, a method called the multicomponent (MC)-MO method has been successfully applied to a number of systems.[47–48] The MC-MO method is an extension of the nonrelativistic MO method and can handle multiple types of particles in a system, namely, electrons and positrons in this context. The Hamiltonian of this nonrelativistic MC (NRMC)-MO method is given by the following:[48]

$$\begin{aligned} H^{(\text{NRMC})} = & \sum_{pq}^{(+)} h_{pq}^{(\text{ele.NR})} a_p^\dagger a_q + \sum_{pq}^{(-)} h_{pq}^{(\text{pos.NR})} b_p^\dagger b_q \\ & + \frac{1}{2}\sum_{pqrs}^{(+)} \left(\psi_p^{(\text{ele.})}\psi_q^{(\text{ele.})}|\psi_r^{(\text{ele.})}\psi_s^{(\text{ele.})}\right) a_p^\dagger a_r^\dagger a_s a_q \\ & + \frac{1}{2}\sum_{pqrs}^{(-)} \left(\psi_p^{(\text{pos.})}\psi_q^{(\text{pos.})}|\psi_r^{(\text{pos.})}\psi_s^{(\text{pos.})}\right) b_p^\dagger b_r^\dagger b_s b_q \\ & - \sum_{pq}^{(+)}\sum_{rs}^{(-)} \left(\psi_p^{(\text{ele.})}\psi_q^{(\text{ele.})}|\psi_r^{(\text{pos.})}\psi_s^{(\text{pos.})}\right) a_p^\dagger b_r^\dagger b_s a_q \end{aligned} \quad (62)$$

with

$$h_{pq}^{(\text{ele.NR})} = \left\langle \psi_p^{(\text{ele.})} \middle| V^{(\text{ext.})} \middle| \psi_q^{(\text{ele.})} \right\rangle + \left\langle \psi_p^{(\text{ele.})} \middle| -\nabla^2/(2m) \middle| \psi_q^{(\text{ele.})} \right\rangle, \tag{63}$$

$$h_{pq}^{(\text{pos.NR})} = -\left\langle \psi_p^{(\text{pos.})} \middle| V^{(\text{ext.})} \middle| \psi_q^{(\text{pos.})} \right\rangle + \left\langle \psi_p^{(\text{pos.})} \middle| -\nabla^2/(2m) \middle| \psi_q^{(\text{pos.})} \right\rangle, \tag{64}$$

where the sum ranges (+) and (−) run over the nonrelativistic MOs of the electron and positron, respectively, rather than the positive- and negative-energy orbitals in the relativistic case. If the negative-energy solution holes in the QED Hamiltonian correctly represent the positron, then the nonrelativistic limit of the QED Hamiltonian should be identical to $H^{(\text{NRMC})}$, except for the difference in the constant terms. The procedure for obtaining the nonrelativistic limit is to first set

$$\psi_p = \begin{bmatrix} \psi_p^{(\text{ele.})} \\ (\boldsymbol{\sigma} \cdot \mathbf{p}/(2mc))\psi_p^{(\text{ele.})} \end{bmatrix} \tag{65}$$

for the *p*-th spinor with positive energy and

$$\psi_p = \begin{bmatrix} -(\boldsymbol{\sigma} \cdot \mathbf{p}/(2mc))\psi_p^{(\text{pos.})} \\ \psi_p^{(\text{pos.})} \end{bmatrix} \tag{66}$$

for the *p*-th spinor with negative energy, and then take the limit $c \to +\infty$.

The resulting nonrelativistic limits for one-electron Hamiltonian matrix elements are

$$h_{pq} \to h_{pq}^{(\text{ele.NR})} + o(c^{-2}) \tag{67}$$

for the case of $p, q \in (+)$,

$$h_{pq} \to -h_{pq}^{(\text{pos.NR})} - 2mc^2 \delta_{pq} + o(c^{-2}) \tag{68}$$

for the case of $p, q \in (-)$, and $h_{pq} \sim c^{-1}$ for the case of $p \in (+), q \in (-)$ or $p \in (-), q \in (+)$. Note that from the orthonormalization condition for the negative-energy spinors, the following relationship holds:

$$\left\langle \psi_p^{(\text{pos.})} \middle| \psi_q^{(\text{pos.})} \right\rangle = \delta_{pq} - \frac{1}{2mc^2} \left\langle \psi_p^{(\text{pos.})} \middle| -\nabla^2/(2m) \middle| \psi_q^{(\text{pos.})} \right\rangle. \tag{69}$$

For the two-electron integrals, the nonrelativistic limit is obtained by extracting the terms independent of *c*, considering that the inner product between the positive- and negative-energy spinors necessarily yields a term with a negative power of *c*.

The resulting nonrelativistic limit of the QED(MO-CNC) Hamiltonian is

$$H^{\text{QED(MO-CNC)}} \to H^{(\text{NRMC})} + 2mc^2 \times (\text{number of positrons}) + o(c^{-1}). \tag{70}$$

The term proportional to $c^2$ on the rhs can be neglected because it is the relativistic rest energy. Since the negative power term of *c* vanishes in the nonrelativistic limit, the nonrelativistic limit of the QED(MO-CNC) Hamiltonian coincides with Eq. (62).

The nonrelativistic limit of the QED(free-CNC) Hamiltonian is

$$H^{\text{QED(free-CNC)}} \to H^{(\text{NRMC})} - \sum_p^{(-)} \left( h_{pp}^{(\text{pos.NR})} - h_{[pp]}^{(\text{pos.NR})} \right)$$

$$+ \frac{1}{2} \sum_{pq}^{(-)} \left\{ \left( \psi_p^{(\text{pos.})} \psi_p^{(\text{pos.})} \mid \psi_q^{(\text{pos.})} \psi_q^{(\text{pos.})} \right) - \left( \psi_p^{(\text{pos.})} \psi_q^{(\text{pos.})} \mid \psi_q^{(\text{pos.})} \psi_p^{(\text{pos.})} \right) \right\}$$

$$- \sum_{pq}^{(-)} \left\{ \left( \psi_p^{(\text{pos.})} \psi_p^{(\text{pos.})} \mid \psi_{[q]}^{(\text{pos.})} \psi_{[q]}^{(\text{pos.})} \right) - \left( \psi_p^{(\text{pos.})} \psi_{[q]}^{(\text{pos.})} \mid \psi_{[q]}^{(\text{pos.})} \psi_p^{(\text{pos.})} \right) \right\}$$

$$+ \frac{1}{2} \sum_{pq}^{(-)} \left\{ \left( \psi_{[p]}^{(\text{pos.})} \psi_{[p]}^{(\text{pos.})} \mid \psi_{[q]}^{(\text{pos.})} \psi_{[q]}^{(\text{pos.})} \right) - \left( \psi_{[p]}^{(\text{pos.})} \psi_{[q]}^{(\text{pos.})} \mid \psi_{[q]}^{(\text{pos.})} \psi_{[p]}^{(\text{pos.})} \right) \right\}$$

$$+ \sum_{pq}^{(+)} \sum_r^{(-)} \left\{ \left( \psi_p^{(\text{ele.})} \psi_q^{(\text{ele.})} \mid \psi_r^{(\text{pos.})} \psi_r^{(\text{pos.})} \right) - \left( \psi_p^{(\text{ele.})} \psi_q^{(\text{ele.})} \mid \psi_{[r]}^{(\text{pos.})} \psi_{[r]}^{(\text{pos.})} \right) \right\} a_p^\dagger a_q$$

$$+ \sum_{pq}^{(-)} \sum_r^{(-)} \left\{ \left( \psi_p^{(\text{pos.})} \psi_q^{(\text{pos.})} \mid \psi_r^{(\text{pos.})} \psi_r^{(\text{pos.})} \right) - \left( \psi_p^{(\text{pos.})} \psi_q^{(\text{pos.})} \mid \psi_{[r]}^{(\text{pos.})} \psi_{[r]}^{(\text{pos.})} \right) \right.$$

$$\left. - \left( \psi_p^{(\text{pos.})} \psi_r^{(\text{pos.})} \mid \psi_r^{(\text{pos.})} \psi_q^{(\text{pos.})} \right) + \left( \psi_p^{(\text{pos.})} \psi_{[r]}^{(\text{pos.})} \mid \psi_{[r]}^{(\text{pos.})} \psi_q^{(\text{pos.})} \right) \right\} b_p^\dagger b_q$$

$$+ 2mc^2 \times (\text{number of positrons}) + o(c^{-1}). \tag{71}$$

This Hamiltonian contains operators in terms that do not depend on $c$; hence it does not coincide with $H^{(\text{NRMC})}$ in the nonrelativistic limit. Similarly, the QED(MO-CCC), QED(free-CCC), QED(MO-cC), and QED(free-cC) Hamiltonians in the nonrelativistic limit are as follows:

$$H^{\text{QED(MO-CCC)}} \to H^{(\text{NRMC})}$$

$$- \frac{1}{2} \sum_{pq}^{(+)} \sum_r^{(+)} \left\{ \left( \psi_p^{(\text{ele.})} \psi_q^{(\text{ele.})} \mid \psi_r^{(\text{ele.})} \psi_r^{(\text{ele.})} \right) - \left( \psi_p^{(\text{ele.})} \psi_r^{(\text{ele.})} \mid \psi_r^{(\text{ele.})} \psi_q^{(\text{ele.})} \right) \right\} a_p^\dagger a_q$$

$$+ \frac{1}{2} \sum_{pq}^{(+)} \sum_r^{(-)} \left( \psi_p^{(\text{ele.})} \psi_q^{(\text{ele.})} \mid \psi_r^{(\text{pos.})} \psi_r^{(\text{pos.})} \right) a_p^\dagger a_q$$

$$+ \frac{1}{2} \sum_{pq}^{(-)} \sum_r^{(+)} \left( \psi_p^{(\text{pos.})} \psi_q^{(\text{pos.})} \mid \psi_r^{(\text{ele.})} \psi_r^{(\text{ele.})} \right) b_q^\dagger b_p$$

$$- \frac{1}{2} \sum_{pq}^{(-)} \sum_r^{(-)} \left\{ \left( \psi_p^{(\text{pos.})} \psi_q^{(\text{pos.})} \mid \psi_r^{(\text{pos.})} \psi_r^{(\text{pos.})} \right) - \left( \psi_p^{(\text{pos.})} \psi_r^{(\text{pos.})} \mid \psi_r^{(\text{pos.})} \psi_q^{(\text{pos.})} \right) \right\} b_q^\dagger b_p$$

$$+ 2mc^2 \times (\text{number of positrons}) + o(c^{-1}), \tag{72}$$

$$
\begin{aligned}
H^{\text{QED(free-CCC)}} \to{} & H^{(\text{NRMC})} - \sum_p^{(-)} \left( h_{pp}^{(\text{pos.NR})} - h_{[pp]}^{(\text{pos.NR})} \right) \\
& + \frac{1}{2} \sum_{pq}^{(-)} \left\{ \left( \psi_p^{(\text{pos.})} \psi_p^{(\text{pos.})} \,|\, \psi_q^{(\text{pos.})} \psi_q^{(\text{pos.})} \right) - \left( \psi_p^{(\text{pos.})} \psi_q^{(\text{pos.})} \,|\, \psi_q^{(\text{pos.})} \psi_p^{(\text{pos.})} \right) \right\} \\
& - \frac{1}{2} \sum_{pq}^{(-)} \left\{ \left( \psi_p^{(\text{pos.})} \psi_p^{(\text{pos.})} \,|\, \psi_{[q]}^{(\text{pos.})} \psi_{[q]}^{(\text{pos.})} \right) - \left( \psi_p^{(\text{pos.})} \psi_{[q]}^{(\text{pos.})} \,|\, \psi_{[q]}^{(\text{pos.})} \psi_p^{(\text{pos.})} \right) \right\} \\
& + \frac{1}{2} \sum_p^{(-)} \sum_q^{(+)} \left\{ \left( \psi_{[p]}^{(\text{pos.})} \psi_{[p]}^{(\text{pos.})} \,|\, \psi_{[q]}^{(\text{ele.})} \psi_{[q]}^{(\text{ele.})} \right) - \left( \psi_p^{(\text{pos.})} \psi_p^{(\text{pos.})} \,|\, \psi_{[q]}^{(\text{ele.})} \psi_{[q]}^{(\text{ele.})} \right) \right\} \\
& - \frac{1}{2} \sum_{pq}^{(+)} \sum_r^{(+)} \left\{ \left( \psi_p^{(\text{ele.})} \psi_q^{(\text{ele.})} \,|\, \psi_{[r]}^{(\text{ele.})} \psi_{[r]}^{(\text{ele.})} \right) - \left( \psi_p^{(\text{ele.})} \psi_{[r]}^{(\text{ele.})} \,|\, \psi_{[r]}^{(\text{ele.})} \psi_q^{(\text{ele.})} \right) \right\} a_p^\dagger a_q \\
& + \sum_{pq}^{(+)} \sum_r^{(-)} \left\{ \left( \psi_p^{(\text{ele.})} \psi_q^{(\text{ele.})} \,|\, \psi_r^{(\text{pos.})} \psi_r^{(\text{pos.})} \right) - \frac{1}{2} \left( \psi_p^{(\text{ele.})} \psi_q^{(\text{ele.})} \,|\, \psi_{[r]}^{(\text{pos.})} \psi_{[r]}^{(\text{pos.})} \right) \right\} a_p^\dagger a_q \\
& + \frac{1}{2} \sum_{pq}^{(-)} \sum_r^{(+)} \left( \psi_p^{(\text{pos.})} \psi_q^{(\text{pos.})} \,|\, \psi_{[r]}^{(\text{ele.})} \psi_{[r]}^{(\text{ele.})} \right) b_p^\dagger b_q \\
& - \sum_{pq}^{(-)} \sum_r^{(-)} \left\{ \left( \psi_p^{(\text{pos.})} \psi_q^{(\text{pos.})} \,|\, \psi_r^{(\text{pos.})} \psi_r^{(\text{pos.})} \right) - \left( \psi_p^{(\text{pos.})} \psi_r^{(\text{pos.})} \,|\, \psi_r^{(\text{pos.})} \psi_q^{(\text{pos.})} \right) \right\} b_p^\dagger b_q \\
& + \frac{1}{2} \sum_{pq}^{(-)} \sum_r^{(-)} \left\{ \left( \psi_p^{(\text{pos.})} \psi_q^{(\text{pos.})} \,|\, \psi_{[r]}^{(\text{pos.})} \psi_{[r]}^{(\text{pos.})} \right) - \left( \psi_p^{(\text{pos.})} \psi_{[r]}^{(\text{pos.})} \,|\, \psi_{[r]}^{(\text{pos.})} \psi_q^{(\text{pos.})} \right) \right\} b_p^\dagger b_q \\
& + 2mc^2 \times (\text{number of positrons}) + o(c^{-1}),
\end{aligned}
\tag{73}
$$

$$
\begin{aligned}
H^{\text{QED(MO-cC)}} \to{} & H^{(\text{NRMC})} \\
& + \sum_{pq}^{(+)} \sum_r^{(-)} \left( \psi_p^{(\text{ele.})} \psi_q^{(\text{ele.})} \,|\, \psi_r^{(\text{pos.})} \psi_r^{(\text{pos.})} \right) a_p^\dagger a_q \\
& - \sum_{pq}^{(-)} \sum_r^{(-)} \left\{ \left( \psi_p^{(\text{pos.})} \psi_q^{(\text{pos.})} \,|\, \psi_r^{(\text{pos.})} \psi_r^{(\text{pos.})} \right) - \left( \psi_p^{(\text{pos.})} \psi_r^{(\text{pos.})} \,|\, \psi_r^{(\text{pos.})} \psi_q^{(\text{pos.})} \right) \right\} b_p^\dagger b_q \\
& + 2mc^2 \times (\text{number of positrons}) + o(c^{-1}),
\end{aligned}
\tag{74}
$$

$$H^{\text{QED(free-cC)}} \to H^{(\text{NRMC})} - \sum_p^{(-)} \left( h_{pp}^{(\text{pos.NR})} - h_{[pp]}^{(\text{pos.NR})} \right)$$

$$+ \frac{1}{2} \sum_{pq}^{(-)} \left\{ \left( \psi_p^{(\text{pos.})} \psi_p^{(\text{pos.})} | \psi_q^{(\text{pos.})} \psi_q^{(\text{pos.})} \right) - \left( \psi_p^{(\text{pos.})} \psi_q^{(\text{pos.})} | \psi_q^{(\text{pos.})} \psi_p^{(\text{pos.})} \right) \right\}$$

$$- \frac{1}{2} \sum_{pq}^{(-)} \left\{ \left( \psi_{[p]}^{(\text{pos.})} \psi_{[p]}^{(\text{pos.})} | \psi_{[q]}^{(\text{pos.})} \psi_{[q]}^{(\text{pos.})} \right) - \left( \psi_{[p]}^{(\text{pos.})} \psi_{[q]}^{(\text{pos.})} | \psi_{[q]}^{(\text{pos.})} \psi_{[p]}^{(\text{pos.})} \right) \right\} \quad (75)$$

$$+ \sum_{pq}^{(+)} \sum_r^{(-)} \left( \psi_p^{(\text{ele.})} \psi_q^{(\text{ele.})} | \psi_r^{(\text{pos.})} \psi_r^{(\text{pos.})} \right) a_p^\dagger a_q$$

$$- \sum_{pq}^{(-)} \sum_r^{(-)} \left\{ \left( \psi_p^{(\text{pos.})} \psi_q^{(\text{pos.})} | \psi_r^{(\text{pos.})} \psi_r^{(\text{pos.})} \right) - \left( \psi_p^{(\text{pos.})} \psi_r^{(\text{pos.})} | \psi_r^{(\text{pos.})} \psi_q^{(\text{pos.})} \right) \right\} b_p^\dagger b_q$$

$$+ 2mc^2 \times (\text{number of positrons}) + o\left( c^{-1} \right),$$

which do not coincide with $H^{(\text{NRMC})}$ in the nonrelativistic limit. Furthermore, QED(Furry-CNC), QED(Furry-CCC), and QED(Furry-cC), obtained from the replacement of parentheses (the parentheses $[\ ]$ replaced by $[\![\ ]\!]$) in Eqs. (71), (73), and (75), respectively, also do not coincide with $H^{(\text{NRMC})}$ in the nonrelativistic limit.

Thus, only the QED(MO-CNC) Hamiltonian properly describes the electron–positron interactions in the nonrelativistic limit. The properties of the Hamiltonians with respect to the three criteria, namely, the orbital rotation invariance, CT invariance, and nonrelativistic limit are summarized in Table 1. Hereafter, we refer to the QED(MO-CNC) Hamiltonian simply as the QED Hamiltonian, unless otherwise noted.

**TABLE I.** Properties of various DC Hamiltonians.

|  | Orbital rotation invariance | CT invariance | non-relativistic limit |
|---|:---:|:---:|:---:|
| VPA | ✓ | ✗ | N/A |
| NVPA | ✗ | ✗ | N/A |
| QED(MO-CNC) | ✗ | ✓ | ✓ |
| QED(free-CNC) | ✓ | ✓ | ✗ |
| QED(Furry-CNC) | ✓ | ✓ | ✗ |
| QED(MO-CCC) | ✗ | ✓ | ✗ |
| QED(free-CCC) | ✓ | ✓ | ✗ |
| QED(Furry-CCC) | ✓ | ✓ | ✗ |
| QED(MO-cC) | ✗ | ✗ | ✗ |
| QED(free-cC) | ✓ | ✗ | ✗ |
| QED(Furry-cC) | ✓ | ✗ | ✗ |

## 4. Total energy expression for QED Hamiltonian

In the discussions in Subsecs. 1–3 of Sec. II C, we have shown that the QED(MO-CNC) Hamiltonian is a candidate for relativistic Hamiltonians that can describe many-body systems including both electrons and positrons. However, in addition, we need to define the total energy expressions for the QED Hamiltonian. Because the QED Hamiltonian does not conserve the number of particles, the number of particles fluctuates, even in vacuum, i.e., the lowest energy state. This indicates that the energy of the vacuum, defined as the expectation value of the QED(MO-CNC) Hamiltonian, may diverge if an infinite number of electron–positron pairs are created, albeit with very low probability. Therefore, to guarantee finite total energy, an offset of the vacuum energy is required: i.e., aligning the vacuum energy with the origin. As such a vacuum, we adopt the Furry vacuum $\left|0^{(\text{Furry})}\right\rangle$ constructed from the eigen orbital set of a one-electron Hamiltonian. Thus, we define the following as the expression of total energy of state $|\Psi\rangle$

$$E^{\text{QED}} = \left\langle \Psi \left| H^{\text{QED}} \right| \Psi \right\rangle - \left\langle 0^{(\text{Furry})} \left| H^{\text{QED}} \right| 0^{(\text{Furry})} \right\rangle, \qquad (76)$$

where the first term represents the un-offset total energy and the second term represents the total energy of the Furry vacuum—these we refer to as the main term and the counter term, respectively. Since electron–positron pair creations are described as excitations from a negative-energy orbital to a positive-energy orbital, the intensity of the divergence depends on the calculation levels of electron correlation. Thus, to properly circumvent the total energy, the level of electron correlation in the counter term should be identical to the main term. For example, at the Dirac–Hartree–Fock (DHF) level, the Furry vacuum state is defined in Eq. (16). At this level, the counter term is zero and there is no need to consider it. In another case, at the CI level, the Furry vacuum state is defined by

$$\left|0^{(\text{Furry})}_{\text{CI}}\right\rangle = \left( \breve{C}_0 + \sum_i^{(+)} \sum_j^{(-)} \breve{C}_j^i a^\dagger_{[\![i]\!]} a_{[\![j]\!]} + \sum_{ik}^{(+)} \sum_{jl}^{(-)} \breve{C}_{jl}^{ik} a^\dagger_{[\![i]\!]} a_{[\![j]\!]} a^\dagger_{[\![k]\!]} a_{[\![l]\!]} + \cdots \right) \left|0^{(\text{Furry})}_0\right\rangle, \qquad (77)$$

where $\breve{C}_0, \breve{C}_j^i, \breve{C}_{jl}^{ik} \cdots$ are the CI coefficients for the vacuum state, which should be determined so that $E^{\text{CI}}_{\text{counter}} = \left\langle 0^{(\text{Furry})}_{\text{CI}} \left| H^{\text{QED}} \right| 0^{(\text{Furry})}_{\text{CI}} \right\rangle$ is minimized. In this case, this minimum value itself is the counter term. The Hamiltonian in the counter term is also expressed in terms of MOs determined from the one-electron Hamiltonian, which always preserves CT invariance. This counter term means that the energy of the system without real (non-virtual) electrons and positrons, or, in other words, the system with only a nucleus, is zero. Noting that the counter term is the energy of the vacuum and that the orbitals corresponding to MOs in the vacuum are Furry orbitals, the following QED Hamiltonian must be used to evaluate the counter term $\left\langle 0^{(\text{Furry})} \left| H^{\text{QED}} \right| 0^{(\text{Furry})} \right\rangle$:

$$H^{\text{QED(MO-CNC)}} = H^{\text{VPA}} - \sum_p^{(-)} h_{[\![pp]\!]} + \frac{1}{2} \sum_{pq}^{(-)} \left[ ([\![pp]\!]|[\![qq]\!]) - ([\![pq]\!]|[\![qp]\!]) \right]$$
$$- \sum_{pq}^{\text{all}} \sum_r^{(-)} \left[ ([\![pq]\!]|[\![rr]\!]) - ([\![pr]\!]|[\![rq]\!]) \right] a^\dagger_{[\![p]\!]} a_{[\![q]\!]}. \qquad (78)$$

We can also adopt another total energy expression, rather than Eq. (76):

$$E^{\text{QED}} = \left\langle \Psi \left| H^{\text{QED}} \right| \Psi \right\rangle - \left\langle 0^{(\text{MO})} \left| H^{\text{QED}} \right| 0^{(\text{MO})} \right\rangle. \qquad (79)$$

This expression differs from Eq. (76) in that the vacuum in the counter term is defined by the MOs. Eq. (79) can be considered as a generalization of Eq. (113) of Ref. 24, defined using the QED(MO-CCC) Hamiltonian. In contrast to the counter term in Eq. (76), the counter term in Eq. (79) employs the MOs, which is the same orbital set as the main term. Since the different system gives different MOs, the different system gives the different counter term in Eq. (79). This means that even the vacuums corresponding to systems differing only in the numbers of electrons are different from each other. This situation seems somewhat strange. The reason is that if we consider systems that differ only in the number of electrons, the vacuum that we reach by removing electrons from the systems is a unique vacuum, where only the nucleus remains, the energy of which should be $\left\langle 0^{(\text{Furry})} \left| H^{\text{QED}} \right| 0^{(\text{Furry})} \right\rangle$, which is employed in Eq. (76).

We should mention here that expressions of Eqs. (27), (67), and (98) in Ref. 24, or Eqs. (7)–(98) in Ref. 49, look formally similar to the total energies here: Eqs. (76) and (79). However, the terms subtracted from the referenced Hamiltonians are single Slater determinants and cannot remove total energy divergence caused by the generalized electron correlation. Thus, these subtracted terms are not counter terms but the terms that are subtracted as the vacuum expectation value that is discarded in the normal ordering procedure described in Sec. II B.

We have now narrowed down the Hamiltonian to QED(MO-CNC) (Eq. (20)) and obtained a total energy expression (Eq. (76)) necessary to formulate a QED-based MO theory. We next derive the QED-level DHF, CI, MP2, and multireference second order perturbation methods based on Eq. (76).

## D. QED-based MO theory

### 1. QED-based DHF method

In previous discussions, we have shown that the relativistic MO method rederived based on the QED(MO-CNC) Hamiltonian can describe positrons reasonably when using negative-energy solutions. In this subsection, we derive the DHF method, including explicit positrons in the configuration, based on the QED(MO-CNC) Hamiltonian. The ground configuration explicitly including positrons is written using creation operators $a_i^\dagger$, $b_p^\dagger$, as

$$|\Psi_0\rangle = \prod_i^{\text{occ(ele.)}} a_i^\dagger \prod_p^{\text{occ(pos.)}} b_p^\dagger |\text{empty}\rangle . \tag{80}$$

The expectation value of the QED(MO-CNC) Hamiltonian for this configuration is the total energy already shown in Eq. (53) of Ref. 45. The QED-based Fock operator

$$\hat{f} = \hat{h} + \sum_j^{\text{occ(ele.)}} \left(\hat{J}_j - \hat{K}_j\right) - \sum_p^{\text{occ(pos.)}} \left(\hat{J}_p - \hat{K}_p\right) \tag{81}$$

was also obtained by taking the variation of this total energy with respect to the electron orbitals in Ref. 45. Note that this Fock operator is shown in matrix notation in Eq. (60) of Ref. 45.

Following the conventional HF procedure, we introduce the Lagrangian to take into account the orthonormality of the orbitals:

$$L = E - \sum_{ij}^{\text{occ(ele.)}} \left(\langle i | j \rangle - \delta_{ij}\right) \epsilon_{ij} - \sum_{pq}^{\text{occ(pos.)}} \left(\langle p | q \rangle - \delta_{pq}\right)(-\epsilon_{pq})$$
$$- \sum_i^{\text{occ(ele.)}} \sum_p^{\text{occ(pos.)}} \langle i | p \rangle \epsilon_{ip} - \sum_i^{\text{occ(ele.)}} \sum_p^{\text{occ(pos.)}} \langle p | i \rangle (-\epsilon_{pi}) . \tag{82}$$

The signs of the Lagrange multipliers in Eq. (82) were chosen so that proper physical interpretation of the multipliers is assured, which simplifies the final form of the DHF equation. Taking the variations of the $k$-th orbital of electrons for $L$, we have

$$\delta L(\delta k) = \langle \delta k | \hat{f} | k \rangle - \sum_{j}^{\text{occ(ele.)}} \langle \delta k | j \rangle \epsilon_{kj} - \sum_{p}^{\text{occ(pos.)}} \langle \delta k | p \rangle \epsilon_{kp} + \text{c.c.} . \quad (83)$$

Using a similar procedure, we can obtain the results for variations of the $s$-th orbital of positrons

$$\delta L(\delta s) = \langle \delta s | -\hat{f} | s \rangle - \sum_{j}^{\text{occ(ele.)}} \langle \delta s | j \rangle (-\epsilon_{sj}) - \sum_{p}^{\text{occ(pos.)}} \langle \delta s | p \rangle (-\epsilon_{sp}) + \text{c.c.} . \quad (84)$$

Setting the $\delta L$ in Eqs. (83) and (84) to zero, we obtain the following DHF equations:

$$\hat{f} | k \rangle = \sum_{j}^{\text{occ(ele.)}} | j \rangle \epsilon_{kj} + \sum_{p}^{\text{occ(pos.)}} | p \rangle \epsilon_{kp} , \quad (85)$$

$$\hat{f} | s \rangle = \sum_{j}^{\text{occ(ele.)}} | j \rangle \epsilon_{sj} + \sum_{p}^{\text{occ(pos.)}} | p \rangle \epsilon_{sp} . \quad (86)$$

Now, we have two equations, (85) and (86), but they are actually equivalent. Therefore, we only have to solve either of the eigen equations of $\hat{f}$. If the solution is considered as an electron solution, then it is a result of Eq. (85), and if it is considered as a positron solution, then it is a result of Eq. (86). These solutions are readily achieved by using the canonical orbitals to define *orbital energies*, as in the case of non-QED. We can easily distinguish the electron and positron solutions, i.e., positive- and negative-energy solutions, by checking whether the orbital energy is greater or less than $-mc^2$. Canonicalizing and applying the LCAO approximation to Eqs. (85) and (86), we have the Dirac–Hartree–Fock–Roothaan (DHFR) equation

$$\sum_{\nu} F_{\mu\nu} C_{\nu i} = \sum_{\nu} S_{\mu\nu} C_{\nu i} \epsilon_k , \quad (87)$$

where $S_{\mu\nu}$ are the overlap integrals and $F_{\mu\nu}$ are the Fock matrix elements

$$F_{\mu\nu} = \langle \mu | \hat{f} | \nu \rangle = h_{\mu\nu} + \sum_{\lambda\rho} \left( P_{\lambda\rho}^{\text{(ele.)}} - P_{\lambda\rho}^{\text{(pos.)}} \right) \left[ (\mu\nu | \lambda\rho) - (\mu\rho | \lambda\nu) \right] . \quad (88)$$

The total energy expression in the LCAO approximation is written as

$$E = \frac{1}{2} \sum_{\mu\nu} \left( P_{\mu\nu}^{\text{(ele.)}} - P_{\mu\nu}^{\text{(pos.)}} \right) \left( h_{\mu\nu} + F_{\mu\nu} \right) , \quad (89)$$

where $P_{\lambda\rho}^{\text{(ele.)}}$ and $P_{\lambda\rho}^{\text{(pos.)}}$ are electronic and positronic density matrices, respectively,

$$P_{\lambda\rho}^{\text{(ele.)}} = \sum_{j}^{\text{occ(ele.)}} c_{\lambda j}^* c_{\rho j} , \quad P_{\lambda\rho}^{\text{(pos.)}} = \sum_{q}^{\text{occ(pos.)}} c_{\lambda q}^* c_{\rho q} . \quad (90)$$

They are characteristic to the QED-based DHF method. For a system without positrons, the positronic density matrix is equal to the null matrix. In this case, the QED-based DHFR equation becomes identical to the conventional DHFR equation. Nevertheless, note that our QED-based DHF formalism considering positrons and CT invariance can give various information on positrons via negative-energy solutions of the equation. According to the discussion in Subsec. 1 of Sec. II B, the *reduced orbital energy*

$$\tilde{\epsilon}_p = -\epsilon_p - 2mc^2, \quad (91)$$

obtained from the negative orbital energy, represents the effective one-particle energy of the positron, and the form of the small component reflects the symmetry (e.g., $s$-type, $p$-type, and so on) of the orbitals of positrons.

## 2. QED-based CI method

The derivation of the basic equation of the QED-based CI method using the DHF wave function is straightforward. In the QED-based MO theory, the electronic state in nonrelativistic theory, which involves only electrons, is generalized to the state that includes both the electrons and positrons. This generalized electronic state is expressed as a linear combination of configurations (Slater determinants) $|\Psi_J\rangle$ in the CI method,

$$|\Psi_\alpha^{\text{CI}}\rangle = \sum_J C_{J\alpha} |\Psi_J\rangle, \tag{92}$$

where subscript $\alpha$ denotes a generalized electronic state. The generalized configurations $|\Psi_J\rangle$ are conventional electronic configurations or configurations including both electrons and positrons due to pair creations, and the coefficients $C_{J\alpha}$ are obtained as eigenvectors of the CI Hamiltonian matrix,

$$H_{IJ} = \langle \Psi_I | H^{\text{QED}} | \Psi_J \rangle. \tag{93}$$

Note that the QED Hamiltonian mixes the DHF configuration with pair-created/annihilated configurations as well as the electron-excited configurations. From Eqs. (20) and (93), the diagonal and nondiagonal matrix elements of the QED-based CI Hamiltonian matrix are written as

$$H_{JJ} = \sum_p^{\text{occ}J(\text{ele.})} h_{pp} - \sum_r^{\text{occ}J(\text{pos.})} h_{rr} + \frac{1}{2} \sum_{pq}^{\text{occ}J(\text{ele.})} \left[(pp|qq) - (pq|qp)\right]$$
$$- \sum_p^{\text{occ}J(\text{ele.})} \sum_r^{\text{occ}J(\text{pos.})} \left[(pp|rr) - (pr|rp)\right] + \frac{1}{2} \sum_{rs}^{\text{occ}J(\text{pos.})} \left[(rr|ss) - (rs|sr)\right] \tag{94}$$

and

$$H_{IJ} = \sum_{pq}^{\text{all}} \langle \Psi_I | a_p^\dagger a_q | \Psi_J \rangle \left\{ h_{pq} - \sum_r^{(-)} \left[(pq|rr) - (pr|rq)\right] \right\}$$
$$+ \frac{1}{2} \sum_{pqrs}^{\text{all}} \langle \Psi_I | a_p^\dagger a_r^\dagger a_s a_q | \Psi_J \rangle (pq|rs), \tag{95}$$

respectively. Here, occ$J$(ele.) and occ$J$(pos.) in the summations indicate that the orbital labels run over the electronic and positronic occupied orbitals in $|\Psi_J\rangle$, respectively. The CI energies $E_\alpha$ are obtained from the counter term and the eigenvalues of the Hamiltonian matrix,

$$E_\alpha = \sum_{IJ} C_{I\alpha}^* H_{IJ} C_{J\alpha} - \langle 0_{\text{CI}}^{(\text{Furry})} | H^{\text{QED}} | 0_{\text{CI}}^{(\text{Furry})} \rangle. \tag{96}$$

## 3. QED-based MP2 method

The perturbation method differs from the CI method in that the counter term is not straightforward, but is treated perturbatively, as is the main term. The MP2 formula at the QED level has already been derived in Refs. 24, 35, and 36 for general orbitals, as well as the canonical MOs. However, in the perturbation energies presented in these references, Eq. (79) is adopted instead of our counter term Eq. (76)—the case involving real (occupied) positrons, which is the focus of our interest, is not considered. We therefore attempted a formulation based on the QED(MO-CNC) Hamiltonian and the counter term Eq. (76), including the case with positronic occupied orbitals.

The energy up to the second order for the DHF configuration, including the counter term, is given by the following formula

$$E^{\mathrm{MP2}} = E^{\mathrm{DHF}} + E^{(2)} - E^{(2)}_{\mathrm{counter}}, \quad (97)$$

$$E^{(2)} = \sum_{n(\neq \mathrm{DHF})} \frac{\langle \Psi_{\mathrm{DHF}} | V | \Psi_n \rangle \langle \Psi_n | V | \Psi_{\mathrm{DHF}} \rangle}{E^{(0)}_{\mathrm{DHF}} - E^{(0)}_n}, \quad (98)$$

where the counter term in Eq. (76) is expressed by the second-order perturbation energy. Note that the zeroth- plus first-order contributions to the counter term is zero, as discussed for the DHF method.

From the discussion in Subsec. 1 of Sec. II B, we can choose either the particle or hole form to derive a more specific energy expression for MP2. Here, we use the hole form for simplicity in the derivation. As the Brillouin theorem is satisfied also for the QED-based DHF method, the intermediate states $|\Psi_n\rangle$ in Eq. (98) are double excitation configurations in the hole form,

$$|\Psi_n\rangle = a_s^\dagger a_b a_r^\dagger a_a |\Psi_{\mathrm{DHF}}\rangle = |\Psi^{rs}_{ab}\rangle, \quad (99)$$

where these excitations include not only conventional electron excitations, but also pair creations. In the hole form, the partitioning of the Hamiltonian into the unperturbed Hamiltonian and perturbation is given as

$$\begin{aligned}
H^{\mathrm{QED(H)}} &= H_0^{\mathrm{QED(H)}} + V \\
H_0^{\mathrm{QED(H)}} &= \sum_p^{\mathrm{all}} \epsilon_p a_p^\dagger a_p - \sum_p^{(-)} \epsilon_p, \\
V &= H^{\mathrm{QED(H)}} - H_0^{\mathrm{QED(H)}} \\
&= \frac{1}{2} \sum_{pqrs}^{\mathrm{all}} (pq|rs) a_p^\dagger a_r^\dagger a_s a_q + (\text{constant and 1-electron operator}).
\end{aligned} \quad (100)$$

According to this partitioning, the energy denominator and numerator in Eq. (98) are

$$\begin{aligned}
E^{(0)}_{\mathrm{DHF}} - E^{(0)}_n &= \langle \Psi_{\mathrm{DHF}} | H_0^{\mathrm{QED(H)}} | \Psi_{\mathrm{DHF}} \rangle - \langle \Psi_n | H_0^{\mathrm{QED(H)}} | \Psi_n \rangle \\
&= \langle \Psi_{\mathrm{DHF}} | \sum_p^{\mathrm{all}} \epsilon_p a_p^\dagger a_p | \Psi_{\mathrm{DHF}} \rangle - \langle \Psi^{rs}_{ab} | \sum_p^{\mathrm{all}} \epsilon_p a_p^\dagger a_p | \Psi^{rs}_{ab} \rangle \\
&= \epsilon_a + \epsilon_b - \epsilon_r - \epsilon_s
\end{aligned} \quad (101)$$

and

$$\langle \Psi_{\mathrm{DHF}} | V | \Psi^{rs}_{ab} \rangle = (ar|bs) - (as|br), \quad (102)$$

respectively. Thus, the second term on the rhs of Eq. (97) becomes

$$E^{(2)} = \frac{1}{2} \sum_{ab}^{\mathrm{occ}} \sum_{rs}^{\mathrm{vir}} \frac{(ar|bs)(ra|sb)}{\epsilon_a + \epsilon_b - \epsilon_r - \epsilon_s} - \frac{1}{2} \sum_{ab}^{\mathrm{occ}} \sum_{rs}^{\mathrm{vir}} \frac{(ar|bs)(rb|sa)}{\epsilon_a + \epsilon_b - \epsilon_r - \epsilon_s}, \quad (103)$$

which looks like the familiar MP2 formula in the nonrelativistic theory. However, the ranges in the summations are

$$\begin{aligned}
\{\mathrm{occ}\} &= \{\mathrm{occ(ele.)}\} \cup \{\mathrm{vir(pos.)}\} \\
\{\mathrm{vir}\} &= \{\mathrm{vir(ele.)}\} \cup \{\mathrm{occ(pos.)}\}.
\end{aligned} \quad (104)$$

When the system includes no positrons, $\{\mathrm{vir(pos.)}\} = \{(-)\}$ and $\{\mathrm{occ(pos.)}\} = \varnothing$ are satisfied. For comparison, the ranges in the summation for VPA-based MP2 are

$$\begin{aligned}\{\text{occ}\} &= \{\text{occ(ele.)}\} \\ \{\text{vir}\} &= \{\text{vir(ele.)}\} \cup \{(-)\}\end{aligned} \quad (105)$$

Note that negative-energy orbitals are treated differently in QED and VPA. In addition to this difference, the counter term $E^{(2)}_{\text{counter}}$ is also needed in the QED-based MP2. The counter term $E^{(2)}_{\text{counter}}$ is calculated for the vacuum state at the same level as the main term,

$$E^{(2)}_{\text{counter}} = \sum_{n(\neq \text{DHF})} \frac{\langle 0^{(\text{Furry})}_{\text{DHF}} | V^{(\text{Furry})} | \Phi_n \rangle \langle \Phi_n | V^{(\text{Furry})} | 0^{(\text{Furry})}_{\text{DHF}} \rangle}{E^{(0,\text{Furry})}_{\text{DHF}} - E^{(0,\text{Furry})}_n}, \quad (106)$$

where $\left|0^{(\text{Furry})}_{\text{DHF}}\right\rangle$ is given in Eq. (16), and $|\Phi_n\rangle$ and $V^{(\text{Furry})}$ are given by

$$|\Phi_n\rangle = a^\dagger_{[\![u]\!]} a_{[\![d]\!]} a^\dagger_{[\![t]\!]} a_{[\![c]\!]} \left|0^{(\text{Furry})}_{\text{DHF}}\right\rangle = \left|\Phi^{[\![tu]\!]}_{[\![cd]\!]}\right\rangle, \quad (107)$$

and from Eq. (78)

$$\begin{aligned} V^{(\text{Furry})} &= H^{\text{QED}(\text{Furry})} - H^{\text{QED}(\text{Furry})}_0 \\ &= \frac{1}{2} \sum_{pqrs}^{\text{all}} ([\![pq]\!] | [\![rs]\!]) a^\dagger_{[\![p]\!]} a^\dagger_{[\![r]\!]} a_{[\![s]\!]} a_{[\![q]\!]} \\ &\quad + (\text{constant and one-electron operator}), \end{aligned} \quad (108)$$

respectively. Here, the definition of the unperturbed Hamiltonian $H^{\text{QED}(\text{Furry})}_0$ is

$$H^{\text{QED}(\text{Furry})}_0 = \sum_p^{\text{all}} \epsilon_{[\![p]\!]} a^\dagger_{[\![p]\!]} a_{[\![p]\!]} - \sum_p^{(-)} \epsilon_{[\![p]\!]}. \quad (109)$$

From these equations, the numerator in Eq. (106) is calculated using

$$\left\langle 0^{(\text{Furry})}_{\text{DHF}} \left| V^{(\text{Furry})} \right| \Phi^{[\![tu]\!]}_{[\![cd]\!]} \right\rangle = ([\![ct]\!] | [\![du]\!]) - ([\![cu]\!] | [\![dt]\!]), \quad (110)$$

and the denominator in Eq. (106) is calculated as

$$\begin{aligned} E^{(0,\text{Furry})}_{\text{DHF}} - E^{(0,\text{Furry})}_n &= \left\langle 0^{(\text{Furry})}_{\text{DHF}} \left| H^{\text{QED}(\text{Furry})}_0 \right| 0^{(\text{Furry})}_{\text{DHF}} \right\rangle - \left\langle \Psi_n \left| H^{\text{QED}(\text{Furry})}_0 \right| \Psi_n \right\rangle \\ &= \epsilon_{[\![c]\!]} + \epsilon_{[\![d]\!]} - \epsilon_{[\![t]\!]} - \epsilon_{[\![u]\!]}. \end{aligned} \quad (111)$$

Thus, the counter term becomes

$$E^{(2)}_{\text{counter}} = \frac{1}{2} \sum_{cd}^{(-)} \sum_{tu}^{(+)} \frac{([\![ct]\!] | [\![du]\!])([\![tc]\!] | [\![ud]\!])}{\epsilon_{[\![c]\!]} + \epsilon_{[\![d]\!]} - \epsilon_{[\![t]\!]} - \epsilon_{[\![u]\!]}} - \frac{1}{2} \sum_{cd}^{(-)} \sum_{tu}^{(+)} \frac{([\![ct]\!] | [\![du]\!])([\![td]\!] | [\![uc]\!])}{\epsilon_{[\![c]\!]} + \epsilon_{[\![d]\!]} - \epsilon_{[\![t]\!]} - \epsilon_{[\![u]\!]}}. \quad (112)$$

From Eqs. (97), (103), and (112), the final energy expression of the QED-based MP2 is

$$\begin{aligned} E^{\text{MP2}} &= E^{\text{DHF}} + \frac{1}{2} \sum_{ab}^{\text{occ}} \sum_{rs}^{\text{vir}} \frac{(ar|bs)(ra|sb)}{\epsilon_a + \epsilon_b - \epsilon_r - \epsilon_s} - \frac{1}{2} \sum_{ab}^{\text{occ}} \sum_{rs}^{\text{vir}} \frac{(ar|bs)(rb|sa)}{\epsilon_a + \epsilon_b - \epsilon_r - \epsilon_s} \\ &\quad - \frac{1}{2} \sum_{cd}^{(-)} \sum_{tu}^{(+)} \frac{([\![ct]\!] | [\![du]\!])([\![tc]\!] | [\![ud]\!])}{\epsilon_{[\![c]\!]} + \epsilon_{[\![d]\!]} - \epsilon_{[\![t]\!]} - \epsilon_{[\![u]\!]}} + \frac{1}{2} \sum_{cd}^{(-)} \sum_{tu}^{(+)} \frac{([\![ct]\!] | [\![du]\!])([\![td]\!] | [\![uc]\!])}{\epsilon_{[\![c]\!]} + \epsilon_{[\![d]\!]} - \epsilon_{[\![t]\!]} - \epsilon_{[\![u]\!]}}. \end{aligned} \quad (113)$$

### 4. QED-based multiconfigurational quasi-degenerate perturbation theory (MCQDPT)

The same procedure as for the single-reference perturbation theory can also be applied to derive a QED-based multireference perturbation theory. A brief explanation follows.

In the multiconfigurational quasi-degenerate perturbation theory (MCQDPT),[50] the effective Hamiltonian up to the second order is given by

$$H_{\alpha\beta}^{\text{MCQDPT}} = H_{\alpha\beta} + \frac{1}{2}\sum_{I(\neq\text{CAS})}\left\{\frac{\langle\Psi_\alpha|V|\Psi_I\rangle\langle\Psi_I|V|\Psi_\beta\rangle}{E_\beta^{(0)} - E_I^{(0)}} + (\alpha \leftrightarrow \beta)\right\}, \quad (114)$$

where $\Psi_\alpha$ and $\Psi_\beta$ are state functions composed of the determinants in the complete active space (CAS), and $\Psi_I$ are the determinants outside CAS. Here, the perturbation Hamiltonian $V$ of MCQDPT is identical to that given in Eq. (100), but the one-electron term should be explicitly specified,

$$V = \frac{1}{2}\sum_{pqrs}^{\text{all}}(pq|rs)a_p^\dagger a_r^\dagger a_s a_q + \sum_{pq}^{\text{all}}\{h_{pq} - \bar{f}_{pq}\}a_p^\dagger a_q + \text{const.}, \quad (115)$$

where $\bar{f}_{pq}$ is the elements of the modified Fock matrix

$$\bar{f}_{pq} = f_{pq} + \sum_r^{(-)}[(pq|rr) - (pr|rq)]. \quad (116)$$

In brief, by treating negative-energy orbitals as occupied orbitals as well as substituting the Fock matrix with the modified Fock matrix, we can perform MCQDPT calculations for the QED Hamiltonian. If the active orbital space of the main term does not include both positive- and negative-energy orbitals, the counter term of second-order MCQDPT is identical to that of MP2 because the corresponding active space for the counter term does not contain any particles in this case. Due to the very large energy gap between the positive- and negative-energy solutions, there is almost no nondynamical correlation between the electron and positron. If one needs to consider both the electron–electron and positron–positron nondynamical correlations, it is recommended to use, for example, the product space (quasi-complete active space[51,52]) of the electron CAS and the positron CAS instead of the CAS across the positive- and negative-energy orbitals. In the cases of using the product spaces, the counter term of second-order MCQDPT is identical to that of MP2.

## III. NUMERICAL RESULTS AND DISCUSSION

### A. QED-based DHF calculations: Hydride ion and positronium

In the previous section, we generalized the DHF method to a method in QED-based form. An important point of the generalization is that it is extended to handle the systems containing positrons. Therefore, the validity of the generalization is tested by whether the method can describe the properties of positrons. In other words, the orbital shapes and energies of negative-energy solutions must be proper as the occupied or virtual orbitals of a positron. To confirm this point, we performed calculations for the hydride ion and the positronium.

#### 1. Hydride ion

We calculated the hydride ion using the DHF method described in the previous section with $s$, $p$, $d$, $f$, and $g$ orbital-type even-tempered Gaussian basis functions, the exponents of which are given by $\{\zeta = 0.01(1/45)^3(2.0)^{n-1}, n=1,2,\ldots,45\}$. Here, we focus on the negative-energy virtual orbitals rather than the occupied orbitals of electrons. This is because, for systems without positrons, our formulation is identical to the conventional DHF method, which only gives the physical interpretation of negative-energy solutions. In Fig. 1, the reduced energy levels of the negative-energy orbital $\tilde{\epsilon}_k$, which is defined in Eq. (91), are shown. The orbitals with energy $\tilde{\epsilon}_k < 0$ are regarded as bound levels of a positron. The degeneracy pattern of the bounded negative-energy levels is not at all similar to the single-center Coulombic field. Referring to the discussion in Subsec. 2 of Sec. II C, we see the relationship between the symmetry of the original basis functions and that of the negative-energy solutions when they are regarded as positronic orbitals. These orbitals are constructed from the basis functions originally used to describe the $p_{1/2}$ orbital of the electrons, but both correspond to the $s$ orbitals of the positron. The shapes of the small component of the two orbitals with the first- and second-lowest reduced orbital energies are shown in Fig. 2. The curve shown on the left side of Fig. 2 corresponds to the $1s$ orbital of a positron. On the left side of Fig. 2, the function value at the origin is 0.026356 and the slope is 0.024353. The curve on the right side of Fig. 2 corresponds to the $2s$ orbital of a positron. On the right side of Fig. 2, the function value at the origin is 0.015954 and the slope is 0.016820. The fact that the function values and the slopes are almost equal indicates that these orbitals follow the nuclear cusp conditions for $Z = -1$. Thus, the particles corresponding to these orbitals behave repulsively with the central proton and can be interpreted as actually having a positive charge. Since the reduced energies of these orbitals are negative, we can interpret, from the discussion in Subsec. 2 of Sec. II C, that these orbitals are virtual orbitals of positrons bound by the excess negative charge of the hydride ion. Therefore, our reinterpretation in Subsec. 1 of Sec. II D holds no inconsistency regarding the virtual orbitals of positrons.

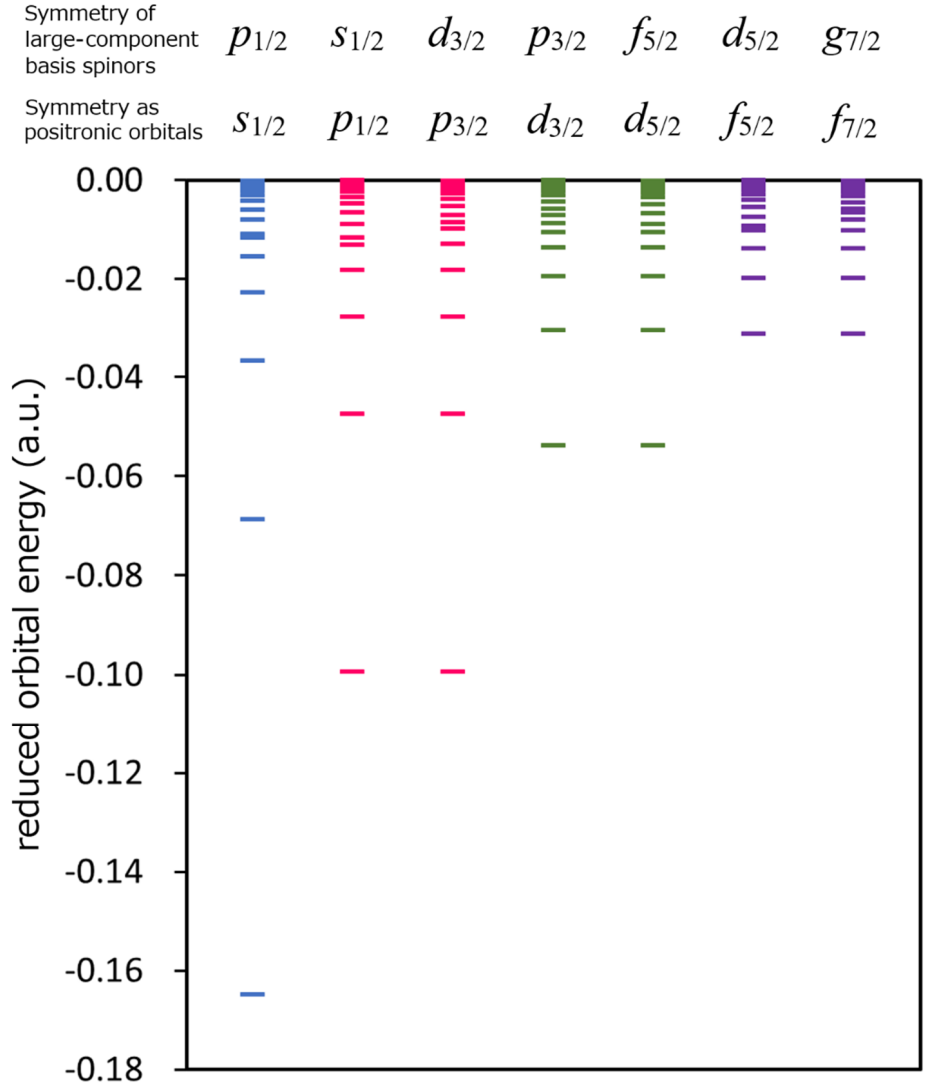

**FIG. 1.** Energy level of negative-energy orbitals for the hydride ion.

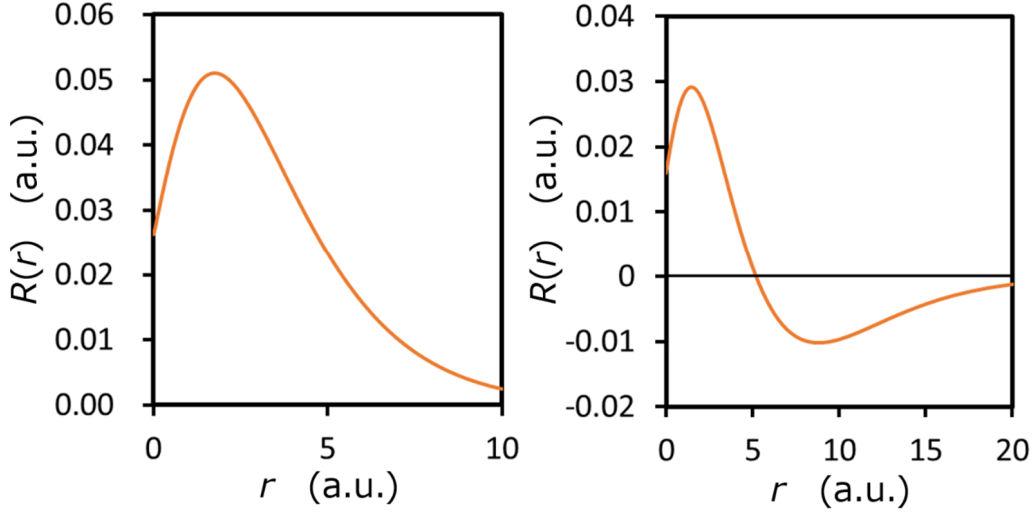

**FIG. 2.** Radial functions of positronic 1s (left) and 2s (right) orbitals for the hydride ion.

2. Positronium

We calculated the singlet configuration of the positronium (para-positronium) using $s$ and $p_{1/2}$ orbital-type even-tempered RKB Gaussian basis functions, the exponents of which are given by $\{\zeta = 0.01(10/45)^3 (2.0)^{n-1}, n = 1, 2, \cdots, 35\}$. In this system, no point charge was placed at the center of the basis functions, and the bound state was formed only by the electron–positron attraction. The radial functions of the calculated (relativistic) orbitals corresponding to the nonrelativistic 1s orbitals of electrons and positrons are shown in Fig. 3 (left). Here, the nonrelativistic 1s orbital of an electron corresponds to the large component of the $s$ orbital with the lowest positive energy, while the nonrelativistic 1s orbital of a positron corresponds to the small component of the $p_{1/2}$ orbital with the highest negative energy. As described in Sec. II B, basis functions employing RKB of the large component of the $s$ orbitals and the small component of the $p_{1/2}$ orbitals have different function forms in the radial direction. Nevertheless, these orbitals are in good agreement, and this indicates that the CT invariance can be recovered by employing a sufficient number of basis functions. The calculated orbital energies corresponding to the 1s orbitals of an electron was –0.162773 a.u., and that of a positron was –37557.562071 a.u., which is equal to –0.162765 a.u. in reduced orbital energy. Thus, the CT invariance of the one-particle energy is also almost preserved to the fifth decimal place.

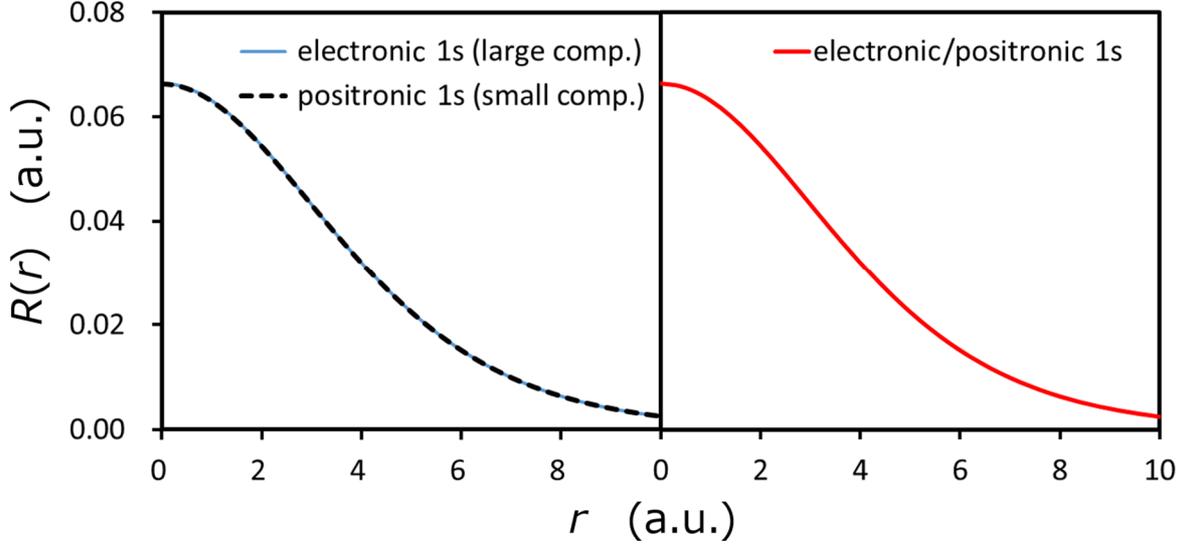

**FIG. 3.** Radial functions of the electronic and positronic 1s orbitals of positronium calculated by the QED-based DHF (left) and the nonrelativistic multicomponent HF (right) methods.

The DHF total energy for this system was –0.100846 a.u., which is considerably higher than the exact nonrelativistic total energy of –0.25 a.u. In general, in precise calculations of the total energy of systems with explicit positrons, the electron–positron correlations are taken into account by highly correlated methods such as the MO method using a Hylleraas-type basis and the diffusion Monte Carlo method. However, since the aim of the numerical calculations in this section is not to calculate positronium compounds accurately, but to show that the negative-energy solution describes the positron, we performed HF calculations for positronium based on the NRMC-MO method. The basis functions used were the $s$-orbital Gaussian basis set with the same exponents as in the DHF calculations above and common to the electron and positron. The 1$s$ orbital radial functions obtained from the calculation were the same for electron and positron, and are shown in Fig. 3 (right). The HF total energy was –0.108513 a.u. The radial function in Fig. 3 (right) is similar to both the large component of electronic 1$s$ spinor and the small component of positronic 1$s$ spinor in Fig. 3 (left). Thus, the results of the DHF calculation and NRMC HF calculation are in good agreement. This indicates that the reinterpretation for negative-energy orbitals shown in Subsec. 1 of Sec. II D is plausible also for the occupied orbitals of positrons.

## B. QED-based MP2 and MCQDPT2 calculations: Helium-like ion

We have expressed our concern about the divergence of the total energy of QED-based electron correlation methods in Subsec. 4 of Sec. II C and our prediction that the problem of this divergence can be solved by a counter term. These concerns and predictions could be shown to exist by numerical calculation of the electron correlation theory in an appropriate system. As a system suitable for the verification, we employed the helium-like ion with $Z$ = 100. As discussed in Subsec. 4 of Sec. II C, the QED-level electron correlation methods involve wave functions consisting of generalized electronic excited configurations, such as pair creations, conventional excitations, and their combinations. Because the number of pair creation configurations becomes very large, even for a system with few electrons, the CI calculations at the QED level are not an easy task. In contrast, the second-order perturbation calculations are readily performed. We therefore performed the second-order perturbation calculations (MP2 and MCQDPT2) as a test of the QED-level electron correlation theory.

To verify whether the MP2 energy converges to a finite value, we observed the change in energy while varying the range of absolute values of the momentum of the orbitals included in the MP2 calculation. For this calculation, we first performed the DHF calculation to obtain the MOs. The basis functions used were $s$-type Gaussian functions with their exponents given by $\{\zeta = 0.01(100/45)^3(2.0)^{n-1}, n = 1, 2, \ldots, 45\}$. The square momenta of the MOs were calculated from the relativistic kinetic energy as

$$\langle p^2 \rangle_i = \left(\langle T_R \rangle_i / c + mc\right)^2 - m^2c^2, \tag{117}$$

where $\langle T_R \rangle_i$ is the relativistic kinetic energy of the $i$-th orbital. We then performed the MP2 calculation by varying the cutoff momentum $p^2$ (i.e., the MOs with $\langle p^2 \rangle_i < p^2$ were included in the orbital summation in the MP2 formula). The results of MP2 calculations are shown in Fig. 4. The horizontal axis indicates the cutoff momentum $p^2$ and the vertical axis indicates the second-order perturbation energy. Note that QED and QED(renormalized) mean $E^{(2)}$ and $E^{(2)} - E^{(2)}_{\text{counter}}$, respectively. With the increase of $p^2$, the perturbation energies of NVPA, VPA, and QED(renormalized) converge to the finite values −0.03176, −0.02687, and −0.03733 a.u., respectively, while the perturbation energies of QED diverge to minus infinity. Thus, in Fig.

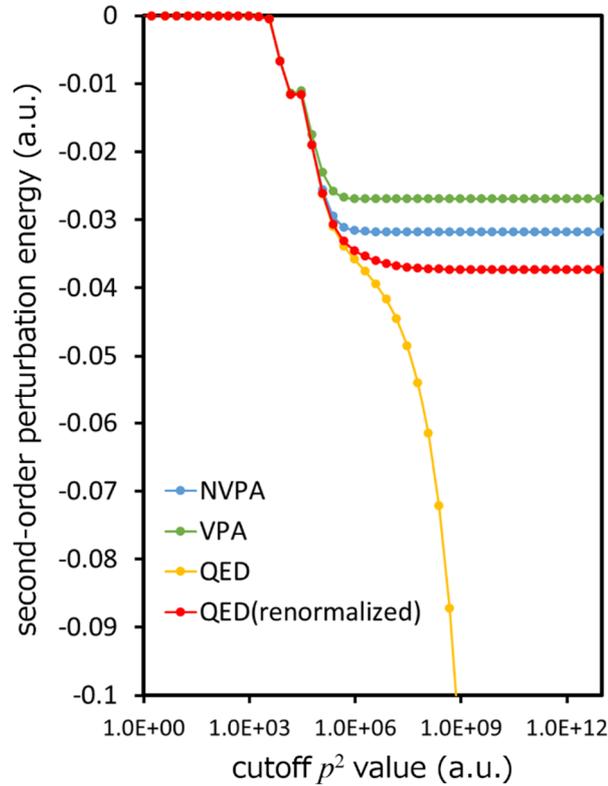

**FIG. 4.** Second-order perturbation energy of MP2 for helium-like ion ($Z$ = 100). The reference DHF total energy (zeroth- and first-order energy) is −11796.85633 a.u.

4, the perturbation energy of VPA is higher than that of NVPA, and that of QED is lower than that of NVPA. This can be understood from the fact that, for second-order perturbations involving negative-energy orbitals, the sign of the denominator of the perturbation is opposite since VPA considers excitation configurations from positive-energy orbitals to negative-energy orbitals and QED considers excitation configurations from negative-energy orbitals to positive-energy orbitals. Here, it should be mentioned that QED(renormalized) is numerically sensitive because it is the difference between two divergent values. Nevertheless, the QED(renormalized) curve in Fig. 4 is a case of smooth convergence. In Fig. 4, the perturbation energy of QED(renormalized) is lower than that of NVPA due to perturbations that do not cancel with the counter term and have a negative value contribution. The excited configurations giving such perturbations are combinations of a single excitation and a single-pair creation.

Subsequently, we performed MCQDPT2 calculations for the helium-like ion with $Z = 100$ to observe the behavior of the excited state energies. The MOs are identical to those in the MP2 calculations. The reference space of MCQDPT2 was the CAS composed of two electrons and the $1s$ and $2s$ orbitals, and the target states were the ground ($S_0$) and three excited ($T_1$, $S_1$, and $S_2$) states, of which the main configurations were the $1s^2$, $1s^1 2s^1$ (triplet), $1s^1 2s^1$ (singlet), and $2s^2$, respectively. The results are shown in Fig. 5 and Table II. In Fig. 5, the horizontal axis is the same as in Fig. 4, and the vertical axis shows the second-order perturbation contributions of the excitation energies. As in the case of the MP2 energy, the perturbation energy of each state diverges with the addition of large momentum configurations. The renormalized total energy of each state is the MCQDPT2 energy minus the counter term energy used in the MP2 calculation, which shows a converged finite value as in the MP2 case. As the counter terms of each state are common, the excitation energies, which are the energy differences from the ground state energy, are the values of the second-order perturbation as it is. The excitation energy thus obtained cancels out the divergence of energy in each state in the same way that the counter terms cancel out each other. In fact, as seen in Fig. 4., the values of the excitation energies for the T$_1$, S$_1$, and S$_2$ states converged to –0.2107, –0.0355, and –0.6178 a.u., respectively, with the increase of $p^2$. Thus, it is shown that the QED-based MP2 and MCQDPT give stable and finite values for the total and excitation energies.

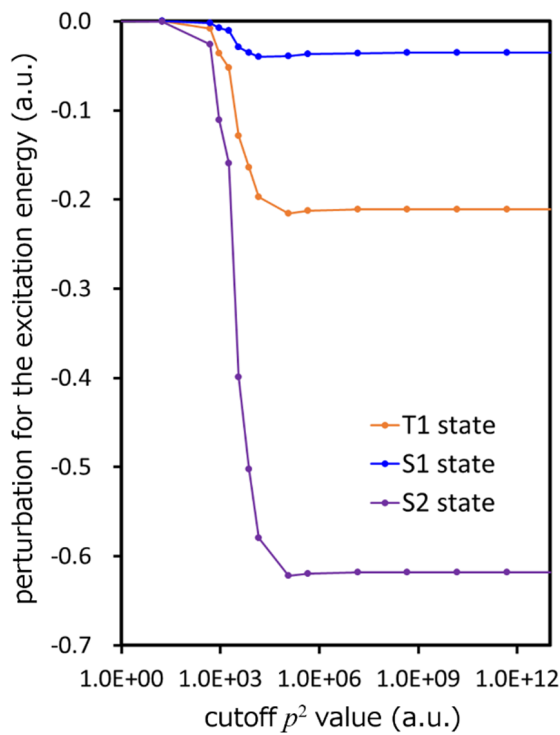

**FIG. 5.** Second-order contribution of MCQDPT2 to excitation energies for the $T_1$, $S_1$, and $S_2$ excited state of a helium-like ion ($Z = 100$).

**TABLE II.** Second-order contributions to excitation energies by MCQDPT. The reference CI (zeroth- plus first-order) excitation energies for the T$_1$, S$_1$, and S$_2$ states are 4334.0179, 4342.2433, and 8719.7062 a.u., respectively, and the S$_0$ state total energy of QED(renormalized), NVPA, and VPA are –11796.89374, –11796.88817, and –11796.88329, respectively.

|     | QED(renormalized) | NVPA    | VPA     |
| --- | ----------------- | ------- | ------- |
| T$_1$ | –0.2107           | –0.2091 | –0.2123 |
| S$_1$ | –0.0355           | –0.0339 | –0.0357 |
| S$_2$ | –0.6178           | –0.6123 | –0.6163 |

### C. QED-based full CI calculations in Fock space: Helium-like ion

In Subsec. 4 of Sec. II B , we showed that the QED Hamiltonian is not orbital rotation invariant and that MOs should be determined so that they give a stationary point of the total energy surface. In this subsection, we confirm the validity of this by means of the QED-based CI method described in Sec. II C. Specifically, we present the results of the QED-based CI method, where the energy surface was created using parameterized MOs, and we discuss the location of the stationary points on the energy surface corresponding to MOs, including the QED-level electron correlation effect.

We employed a helium-like ion with $Z = 100$ for the confirmation. The MOs used in the CI calculations were the spinors and their Kramers pairs in the following form:

$$\chi_i = c_{1i}\begin{bmatrix}\varphi_1 \\ \mathbf{0}_2\end{bmatrix} + \frac{c_{2i}}{\sqrt{1-|\langle\varphi_1|\varphi_2\rangle|^2}}\begin{bmatrix}\varphi_2 - \langle\varphi_1|\varphi_2\rangle\varphi_1 \\ \mathbf{0}_2\end{bmatrix} + c_{3i}\begin{bmatrix}\mathbf{0}_2 \\ \varphi_3\end{bmatrix}, \quad (118)$$
$$(i = 1, 2, 3)$$

where the "AOs" $\varphi_1$, $\varphi_2$, and $\varphi_3$ were

$$\varphi_1 = \begin{bmatrix}(2\alpha_1/\pi)^{3/4}\exp(-\alpha_1 r^2) \\ 0\end{bmatrix},$$

$$\varphi_2 = \begin{bmatrix}(2\alpha_2/\pi)^{3/4}\exp(-\alpha_2 r^2) \\ 0\end{bmatrix}, \quad (119)$$

$$\varphi_3 = \frac{\boldsymbol{\sigma}\cdot\mathbf{p}}{\sqrt{3\alpha_3}}\begin{bmatrix}(2\alpha_3/\pi)^{3/4}\exp(-\alpha_3 r^2) \\ 0\end{bmatrix},$$

with $\alpha_1=100$, $\alpha_2=1000$, and $\alpha_3=100$, and the MO coefficients $c_{\mu i}$ are given by

$$c = \begin{bmatrix} \sin\theta\cos\phi & \cos\theta\cos\phi & -\sin\phi \\ \sin\theta\sin\phi & \cos\theta\sin\phi & \cos\phi \\ \cos\theta & -\sin\theta & 0 \end{bmatrix}. \tag{120}$$

The original ranges of $\theta$ and $\phi$ in spherical coordinates are $0 \leq \theta \leq \pi$ and $0 \leq \phi \leq 2\pi$, respectively, but if their direction is completely opposite, they are the same as the basis set, so their ranges can be changed to $-\pi/2 \leq \theta \leq \pi/2$ and $0 \leq \phi \leq \pi$, respectively. With the parameters $\theta$ and $\phi$, any MOs that can be constructed from the given set of AOs are generated. The two spinors obtained from the coefficients in the first column ($j=1$) and the Kramers pair correspond to positrons, and the remaining four spinors obtained from the coefficients of the second and third columns ($j=2$ and $3$) correspond to electrons.

Using these MOs with various $\theta$ and $\phi$, we performed full CI calculations for two-electron systems. The energy surface of the ground state obtained in this way is shown in Fig. 6. From the contour plot in Fig. 6 (a), we can see that the maximum value exists in the red region and the minimum value in the blue region, and the QED MO can be obtained by these stationary points. In the blue region, the small component of the positive solution is large and the large component of the negative solution is large, i.e., the electron and positron are treated in the opposite way. Thus, the MO that gives the stationary point in the blue region does not match our usual picture. The plausible MO is that corresponding to the stationary point in the red region. The QED MOs obtained in this way are shown in Table III along with the free-particle MOs (MOs obtained by diagonalization of the kinetic energy matrix), Furry MOs (MOs obtained by diagonalization of the core Hamiltonian), and fuzzy MOs (MOs equivalent to the DHF method). Note that the NVPA energy surface has similar features to the QED surface, but it was not possible to define the parameters $\theta$ and $\phi$ uniquely because the maximum points were connected in a curved segment. Furthermore, as VPA has orbital rotation invariance, any MO will give the same energy levels in full CI, and parameters $\theta$ and $\phi$ need not be determined. The QED MOs were close to the Furry MOs and the fuzzy MOs; in particular, the fuzzy MOs were very close to the QED MOs. As mentioned in Sec. II B, proper MOs for a stationary state at the QED level are those giving a stationary value of the total energy surface, and the HF MOs are an approximation to this. In fact, the $\theta$ and $\phi$ values of the QED MOs and the HF MOs were very close, indicating that this approximation was valid. Due to the energy difference of more than $2mc^2$ between the ground and any pair-created configurations, there is almost no static electron correlation between these configurations. This provides a good basis for the approximation. For these reasons, in practice, the HF MOs can be used as MOs at the QED level.

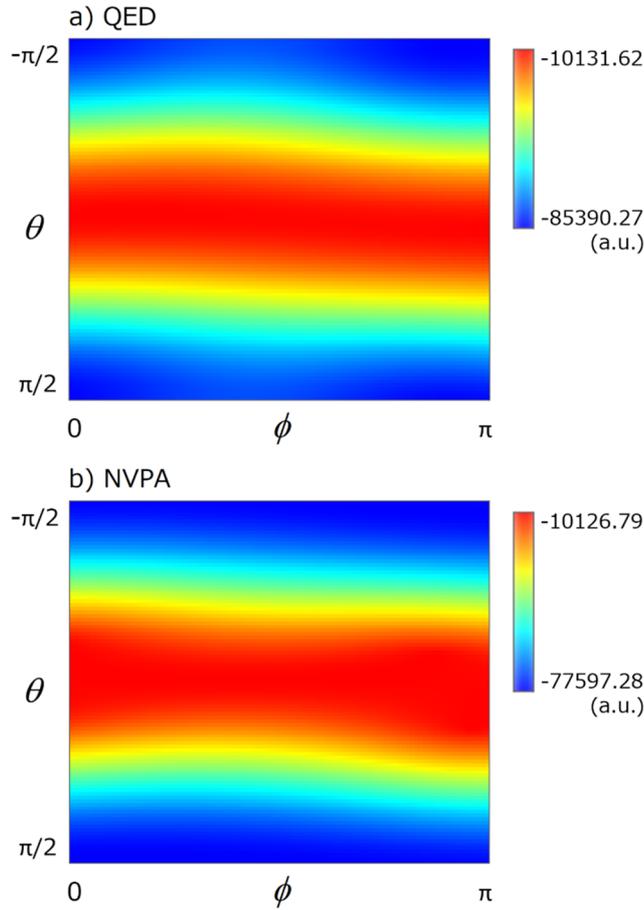

**FIG. 6.** Total energy contour map for the ground state of the helium-like ion ($Z = 100$) toy model—calculated using the full CI method for the whole MO space, using the QED (upper) and NVPA (lower) Hamiltonians.

**TABLE III.** Angle parameters $\theta$ and $\phi$ for the MOs determined by various methods.

|  | $\theta$ | $\phi$ |
|---|---|---|
| Free | –0.0675656 | 0.377358 |
| Furry | –0.0709949 | 0.430620 |
| Fuzzy(Hartree–Fock) | –0.0709650 | 0.430296 |
| Maximum of QED | –0.0709650 | 0.430296 |

## IV. CONCLUSIONS

We have discussed proper forms of the second-quantized relativistic many-body Hamiltonian as an attempt to solve the negative-energy problem in relativistic MO theory. The many-body Hamiltonian that could solve this problem is the so-called QED Hamiltonian, and several variations have been proposed from previous studies. We theoretically investigated the properties of the QED Hamiltonians in terms of the orbital rotation invariance, CT invariance, and nonrelativistic limit. First, we examined the orbital rotation invariance of the QED Hamiltonians. We showed that the three QED Hamiltonians are not orbital rotation invariant. This noninvariance may lead to indefinite eigenvalues. However, we also showed that this problem does not arise if the orbitals are fixed to those proper for a stationary state, and that such proper orbitals are determined as MOs giving a stationary point of the total energy surface. Second, we examined the CT invariance of the QED Hamiltonian. We showed that positron orbitals can be described as negative-energy solutions when CT invariance holds and that the six QED Hamiltonians are CT invariant. Third, we examined the nonrelativistic limit of the QED Hamiltonians. We showed that only the QED(MO-CNC) Hamiltonian has a reasonable nonrelativistic limit consistent with conventional (nonrelativistic) MCMO theory. Among our candidates for the QED Hamiltonians, the only Hamiltonian that meets our requirements is the QED(MO-CNC) Hamiltonian. In addition, we mention the possibility of divergence of the total energy obtained from the QED Hamiltonian, which allows an infinite number of pair creations. To avoid the divergence, we proposed an energy including a counter term, which adjusts the energy origin so that the Furry vacuum energy is zero.

Based on these considerations, we discussed the DHF and electron correlation methods. We showed that the conventional DHF equation can also be regarded as the DHF equation for the QED Hamiltonian, keeping the same form. In addition, we presented CI, single-reference perturbation (MP2), and multireference perturbation (MCQDPT) methods based on the QED Hamiltonian.

In the results of the DHF calculations of the hydride ion, the negative-energy solution can be interpreted consistently as a bound positron solution. It was also found that, although the function forms of the basis functions using RKB lack CT invariance, the CT invariance of the QED Hamiltonian itself can be recovered by using a sufficiently rich basis function set. The results of MP2 calculations for the helium-like ion showed that the total energy of the QED Hamiltonian without the counter term diverges to minus infinity, and the introduction of the counter term leads to a proper finite total energy. The MCQDPT calculations for the helium-like ion also showed that the QED Hamiltonian gives finite excitation energies; in this case, the counter term does not affect the excitation energies since they are the energy differences between the total energies of the states. Finally, from the results of the CI calculations, we suggest that the MOs containing electron correlation effects at the QED level are almost the same as those determined by the DHF method. This indicates that, for practical purposes, MOs of the DHF method can also be used in MO calculations at the QED level. From these considerations and results, it is thus considered appropriate to use the QED(MO-CNC) Hamiltonian as the relativistic Hamiltonian used in constructing the relativistic MO theory.

# ACKNOWLEDGMENTS

This work was supported by the Japan Society for the Promotion of Science (JSPS) Kakenhi, Grant Numbers: JP14J06668 to NI and JP21K04980 to HN.

# AUTHOR DECLARATIONS

Conflict of interest
The authors have no conflicts to disclose.